\documentclass[aps,10pt,prb,nofootinbib,citeautoscript,longbibliography,superscriptaddress,twocolumn]{revtex4-2}
\usepackage{graphicx} 
\usepackage{siunitx}
\usepackage{hyperref}
\usepackage{booktabs}
\hypersetup{
    colorlinks=true,
    linkcolor=blue,
    filecolor=blue,      
    urlcolor=blue,
    citecolor=blue,
}
\makeatletter
\DeclareRobustCommand{\cev}[1]{%
    {\mathpalette\do@cev{#1}}%
}
\newcommand{\do@cev}[2]{%
    \vbox{\offinterlineskip
    \sbox\z@{$\m@th#1 x$}%
    \ialign{##\cr
        \hidewidth\reflectbox{$\m@th#1\vec{}\mkern4mu$}\hidewidth\cr
        \noalign{\kern-\ht\z@}
        $\m@th#1#2$\cr
    }%
    }%
}
\makeatother
\begin{document}
% \preprint{APS/123-QED}

\title{Phase-dependent epitaxy for antimonene growth on silver substrate}  % Force line breaks with \\

%%% authors information
\author{Kai Liu}
\thanks{These authors equally contribute to this work.}

\author{Keke Bai}
\thanks{These authors equally contribute to this work.}

\author{Jing Wang}
\email{jwang@hebtu.edu.cn}

\author{Juntao Song} %
\affiliation{%
    Department of Physics and Hebei Advanced Thin Film Laboratory,
    Hebei Normal University, Shijiazhuang 050024, Hebei, China.
}%

\author{Ying Liu} %

\affiliation{%
    Department of Physics and Hebei Advanced Thin Film Laboratory,
    Hebei Normal University, Shijiazhuang 050024, Hebei, China.
}%
\affiliation{National Key Laboratory for Materials Simulation and Design, Beijing 100083, China}
%%%%%%%%%%%%%%%%%%%%%%%%%%%%%%%%%%%%%%%%%%%%%%%%%%%%%%%%%%%%

\date{\today}% It is always \today, today,
%  but any date may be explicitly specified
\pagenumbering{arabic}

\begin{abstract}

Antimonene is a novel two-dimensional topological semiconductor material with strain driven tunable electronic structure for future electronic and spintronic devices, but the growth of clean antimonene is not fully understood.
In this work, the growth process of antimonene on silver substrate has been studied in detail by using density functional theory and particle swarm optimization algorithms. The results show that, in addition to the experimental reported flat honeycomb and $\beta$-phase antimonene, $\alpha$-phase antimonene was observed to be able to grow on thus substrates, and the phases of antimonene were deeply dependent on the reconstructed supercells and surface alloys. It has been demonstrated that the surface alloys on substrate play an active role in the growth of antimonene.

\end{abstract}

% \begin{keyword}
%     Antimonene\sep Density function theory\sep epitaxial growth\sep Ag(111) surface
% \end{keyword}

% \keywords{Antimonene, Density function theory, epitaxial growth}
% Use showkeys class option if keyword
% display desired

\maketitle

\section{Introduction}\label{sec:intro}

Antimonene, one of the novel mono-elemental class of two-dimensional(2D) material, has been extensively studied from theory to experiment~\cite{aresRecentProgressAntimonene2018,zhangRecentProgress2D2018,wangAntimoneneExperimentalPreparation2019}, since it was first reported by Zhang et al.~\cite{zhangAtomicallyThinArsenene2015} in 2015. As a member of 2D group-V family, the freestanding monolayer antimonene, either puckered ($\alpha$-phase) or buckled ($\beta$-phase) structure, are semiconductors. Compared to $\alpha$-phase antimonene ($\alpha$-Sb), $\beta$-phase antimonene ($\beta$-Sb) with a buckled honeycomb structure \cite{wangAtomicallyThinGroup2015} is more stable. In the past few years, $\beta$-Sb was reported to have a strain-tunable energy gap and high carrier mobility~\cite{zhangAtomicallyThinArsenene2015, wangAtomicallyThinGroup2015, wangElectronicStructureCarrier2015,akturkSinglelayerCrystallinePhases2015,wangThermalTransportProperties2016,zhaoStraindrivenBandInversion2015,kripalaniStrainEngineeringAntimonene2018}. Very recently, by applying a large tensile strain on the lattice, Zhao et al. realized the band-inversion, in which the $\beta$-Sb transferred from trivial semiconductor to nontrivial quantum spin Hall insulator~\cite{zhaoTwodimensionalGraphenelikeXenes2020}. As to $\alpha$-Sb, Cheung et al. found that $\alpha$-Sb can be tuned to a 2D topological insulator under an in-plane anisotropic strain~\cite{cheungSpinOrbitCoupling2016}. All these studies confirm the possibility of antimonene as one of the promising candidates for optoelectronics and spintronic devices.

To obtain monolayer antimonene, scientists have experimented with different preparation methods, including micromechanical exfoliation (MME), liquid phase exfoliation (LPE), and molecular beam epitaxy (MBE). By applying the MME method, which used to obtain graphene \cite{novoselovElectricFieldEffect2004}, Ares et al.~\cite{aresMechanicalIsolationHighly2016} successfully obtained few-layer antimonene, but it's hard to obtain monolayer antimonene. Some other works have reported successes in the preparation of antimonene using LPE method \cite{gibajaFewLayerAntimoneneLiquidPhase2016,wangBandgapTunablePreparationSmooth2018,gibajaLiquidPhaseExfoliation2019}, while the samples obtained were mostly freestanding antimonene. It usually cannot meet the requirements for obtaining topological properties. For the MBE method, it may offer possibility to obtain stress-stretched antimonene. In 2016, Ji et al.~\cite{jiTwodimensionalAntimoneneSingle2016} synthesized high-quality, few-layer $\beta$-phase antimonene ($\beta$-Sb) polygons on mica substrate through van der Waals epitaxy. Since then, there have a number of work successfully epitaxially grown $\beta$-Sb on several types of substrates, such as layered materials, transition metals, and transition metal-oxides. According to our collection, layered material substrates include PdTe$_2$~\cite{wuEpitaxialGrowthAirStability2017}, MoS$_2$~\cite{chenSingleCrystalAntimoneneFilms2018}, graphene \cite{sunVanWaalsEpitaxy2018}, Bi$_2$Se$_3$ \cite{flamminiEvidenceBetaAntimonene2018}, WSe$_2$ \cite{zhangTungstenDiselenideTopgate2020}; transition metal substrates include Ag(111) \cite{maoEpitaxialGrowthHighly2018,shaoEpitaxialGrowthFlat2018,sunRealizationBuckledAntimonene2020}, Cu(111) \cite{niuModulatingEpitaxialAtomic2019}, Au(111) \cite{zhouInterfacialEffectsGrowth2019} surfaces; transition metal-oxide substrates include Cu$_3$O$_2$ \cite{niuLargeScaleSynthesis2020}, sapphire \cite{guDirectGrowthAntimonene2020}.

During the epitaxially grown process of antimonene, the substrate played a very important role. The lower mismatch between the selected substrates and antimonene, the lattice parameters of the grown antimonene were closer to that of freestanding antimonene. For example, the lattice constant of antimonene on PdTe$_2$ was 4.13\AA{} \cite{wuEpitaxialGrowthAirStability2017}, and 4.1 \AA{} \cite{flamminiEvidenceBetaAntimonene2018} on Bi$_2$Se$_3$. The lattice parameter of antimonene grown on transition metals was much larger than that of freestanding antimonene. It was 4.43 \AA{} on Cu(111) surface \cite{niuModulatingEpitaxialAtomic2019} and 5.0 \AA{} on Ag(111) surface \cite{maoEpitaxialGrowthHighly2018,shaoEpitaxialGrowthFlat2018}. Thus, transition metals may be good substrates for epitaxial growth of antimonene with stress tensile.

The Density Functional Theory (DFT) simulations play an increasingly important role in the experimental synthesis and analysis of two-dimensional (2D) materials these years. The structural analysis of 2D materials is one of the main functions.
In this paper, the structural analysis of antimonene on Ag(111) substrate, on which antimonene could have about 20\% stretching of lattice, were carried out with DFT calculations based on the following fundamental questions: (1) How do the structures evolve with increasing Sb coverage? (2) What is the most stable phase of antimonene on substrates?
Our results suggested that the antimonene phase was strongly influenced by the structure of the substrate and reconstruction supercell.

\section{Computation methods}\label{sec:methods}

To find the most stable configuration of antimonene on Ag(111) substrate, we have selected different supercells of Ag(111) surface, and performed surface reconstruction search under different Sb coverage using particle swarm optimization(PSO) algorithm, which is well-designed in the software package CALYPSO \cite{wangCrystalStructurePrediction2010, wangCALYPSOMethodCrystal2012, luSelfassembledUltrathinNanotubes2014}. Our calculations were based on the fact that antimony atoms were in full contract with the substrate and were uniformly distributed, so that the evolution of the most stable structures at different coverage were theoretically able to describe the continuous deposition of Sb-Ag(111) system. The coverage of antimony ($C_{\text{Sb}}$) on the substrate was the key parameter in this work. Usually, we describe it in terms of $C_\text{Sb} = N_\text{Sb} / N_\text{Sb-sheet}$. In this equation, the $N_\text{Sb-sheet}$ would not be a constant for different phases of substrate with same reconstructed area. Therefore, for ease to describe $C_{\text{Sb}}$, we defined it as the following equation:
\begin{equation}
    C_{\text{Sb}} = \frac{N_{\text{Sb}}}{N_{\text{Sub}}}, \label{eq:Csb}
\end{equation}
where, $N_{\text{Sb}}$ is the number of Sb atoms adsorbed on substrate, and $N_{\text{Sub}}$ is the number of atoms in a monolayer of the substrate. For supercells in type of $\mathcal{A}$ ($\sqrt{3}\times\sqrt{3}$), $\mathcal{B}$ ($3\times\sqrt{3}$) and $\mathcal{C}$ ($3\times 3$) of Ag(111), $N_{\text{Sub}}$ = 3, 6 and 9, respectively.

The formation energy of Sb adsorbed onto substrate, for measuring structural stability, is defined by formula:
\begin{equation}
    E_{\text{form}} = \frac{1}{N_{\text{Sb}}}\left( E_{\text{tot}} - E_{\text{sub}} \right) - E_{\text{bulk}}^{\text{Sb}}, \label{eq:eform}
\end{equation}
where, $E_{\text{tot}}$ is the total energy of the substrate with Sb atoms adsorbed, $E_{\text{sub}}$ is the total energy of the substrate, $E_{\text{bulk}}^{\text{Sb}}$ is the bulk total energy per atom of bulk Sb (space group: R-3m, No.166). Under this definition, the lower the formation energy, the more stable the structure, and a negative value of energy indicates that growth is prone to occur.

To our knowledge, the formation of surface alloy on substrate is crucial for the epitaxial growth of two-dimensional materials. It has been reported that when being deposited on the Ag(111) surface, Sb atoms react with the surface and form Ag$_2$Sb surface alloy, which is combined by a monolayer of a $1\times 1$ Ag$_2$Sb matching with a $\sqrt{3}\times\sqrt{3}$ R \ang{30} supercell of Ag(111)\cite{oppoTheoryAdsorptionSurfactant1993, soaresStructureDeterminationAg2000,quinnStructuralStudyAdsorption2002}. Therefore, a transition state search calculation was done to verify the possibility of the existence of surface alloy and to identify their structures. After confirmed the surface alloy structure, we extended these types of supercells with and without surface alloy. To facilitate the description, we ignored the intermediate alloy layer and all reconstructed structure labels are based on Ag(111).

All PSO calculations were energetically converged with a population size of 20 and a max step of 10 (for saving compute resources, 5 steps for N$_\text{Sb}$ = 1 -- 4, 10 or 15 steps for N$_\text{Sb}$ $\geq$ 5) to meet global optimization required. Then, the possible stable surface structures with lower entropy and different structural symmetry were collected to do the next step of high precision structure optimization. After all DFT calculations done, the energetically favorable structures of all coverages were obtained to characterize the most probable structural evolution path.

All DFT calculations were carried out as implemented in the Vienna ab-initio simulation package (VASP) \cite{kresseInitioMoleculardynamicsSimulation1994,kresseEfficiencyAbinitioTotal1996,kresseEfficientIterativeSchemes1996}. The electron-ion interaction was described by the projector augmented-wave potentials \cite{blochlProjectorAugmentedwaveMethod1994}, and the exchange-correlation functional was given by generalized gradient approximation parametrized by Perdew, Burke, and Ernzerhof \cite{perdewAtomsMoleculesSolids1992,perdewGeneralizedGradientApproximation1996}. The van der Waals corrections were treated by the semi-empirical DFT-D3 method \cite{grimmeConsistentAccurateInitio2010} in surface reconstruction calculations. The cutoff energy were set to 400 eV for all calculations. The vacuum layer was set to be at least 15 \AA{} to eliminate the interaction between the layers. All structures were fully relaxed until the force on each atom was less than 0.02 eV/\AA{}, and the energy convergence criterion was met to $10^{-8}$ eV. The substrates were built on the basis of the bulk structure of FCC Ag, and the Ag bulk lattice constant obtained by structural optimization was 4.137 \AA, which was well converged in VASP. The total number of layers in the substrate were set to 3 (4 for substrate with surface alloy), and the atom positions in the bottom layer were fixed during the geometry optimization.

\section{Results and discussion}\label{sec:results}

\subsection{Antimonene on pure Ag(111) surface}\label{sec:res:1}

\begin{figure*}[htbp]
    \centering
    \includegraphics[width=\linewidth]{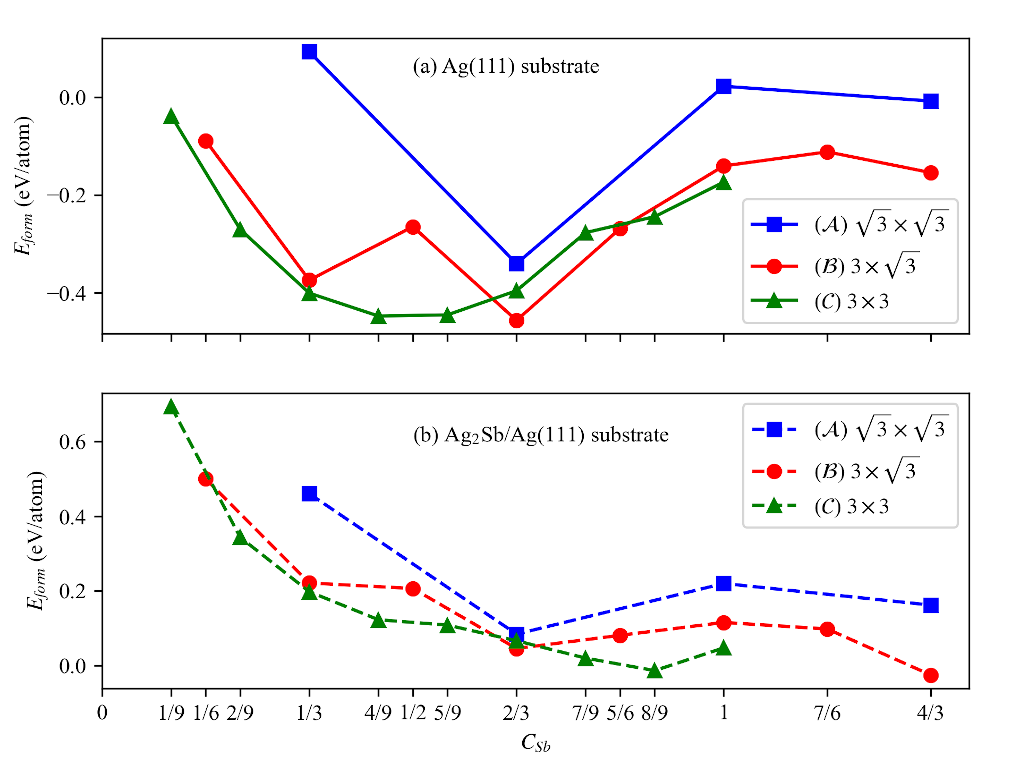}
    \caption{The formation energies of Sb atoms adsorbed on different types of pure Ag(111)  and Ag$_2$Sb/Ag(111) substrates.
        (a) and (b) show the evolution curves of the formation energy, where blue,
        red and green represent $\sqrt{3}\times\sqrt{3}$, $3\times\sqrt{3}$ and $3\times{}3$
        superstructure, solid and dashed lines represent Ag(111) and Ag$_2$Sb/Ag(111) substrates, respectively.}
    \label{fig:eform}
\end{figure*}

\begin{figure*}[htbp]
    \centering
    \includegraphics[width=0.99\linewidth]{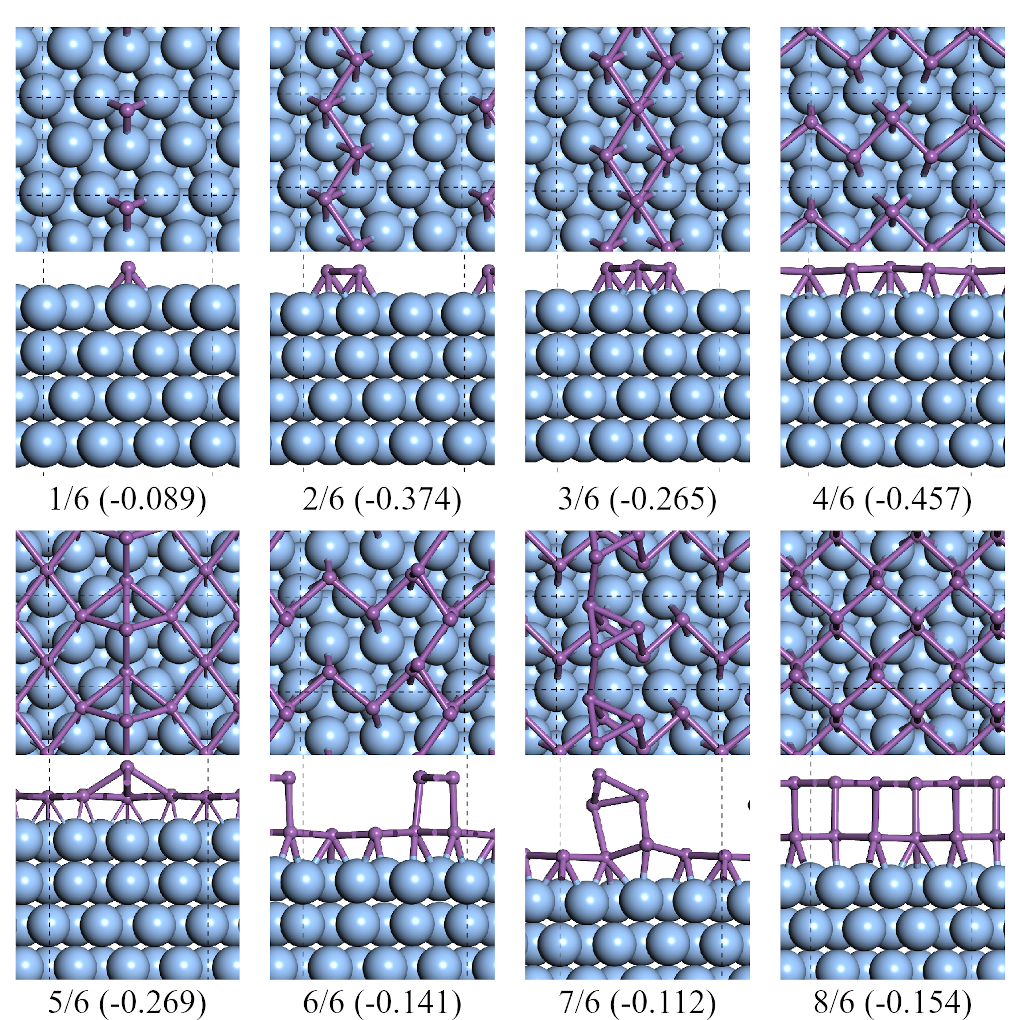}
    \caption{
        Top and side views of the lowest formation energy favorable configurations of antimony adsorbed on $3 \times \sqrt{3}$ Ag(111) surface.
    }
    \label{fig:struc-3r3}
\end{figure*}

In this section, we have carried out the structural search for the most stable configuration with different coverage of Sb atoms on pure Ag(111) substrate.
In Fig.~\ref{fig:eform}(a), we depicted the variation of formation energies with $C_\text{Sb}$ on Ag(111) substrate. The configuration with the lowest formation energy is at $C_{\text{Sb}} = 2/3$ on $\mathcal{B}$-type Ag(111) substrate, in which Sb atoms are arranged in a zigzag shape on the substrate, which is consistent with the half layer of $\alpha$-Sb, see Fig.~\ref{fig:struc-3r3} (4/6).

% \begin{figure}[htbp]
%     \centering
%     \includegraphics[width=\linewidth]{img/eform.png}
%     \caption{The formation energies of Sb atoms adsorbed on different types of pure Ag(111)  and Ag$_2$Sb/Ag(111) substrates.
%         (a) and (b) show the evolution curves of the formation energy, where blue,
%         red and green represent $\sqrt{3}\times\sqrt{3}$, $3\times\sqrt{3}$ and $3\times{}3$
%         superstructure, solid and dashed lines represent Ag(111) and Ag$_2$Sb/Ag(111) substrates, respectively.}
%     \label{fig:eform}
% \end{figure}

\begin{figure*}[htbp]
    \centering
    \includegraphics[width=0.99\linewidth]{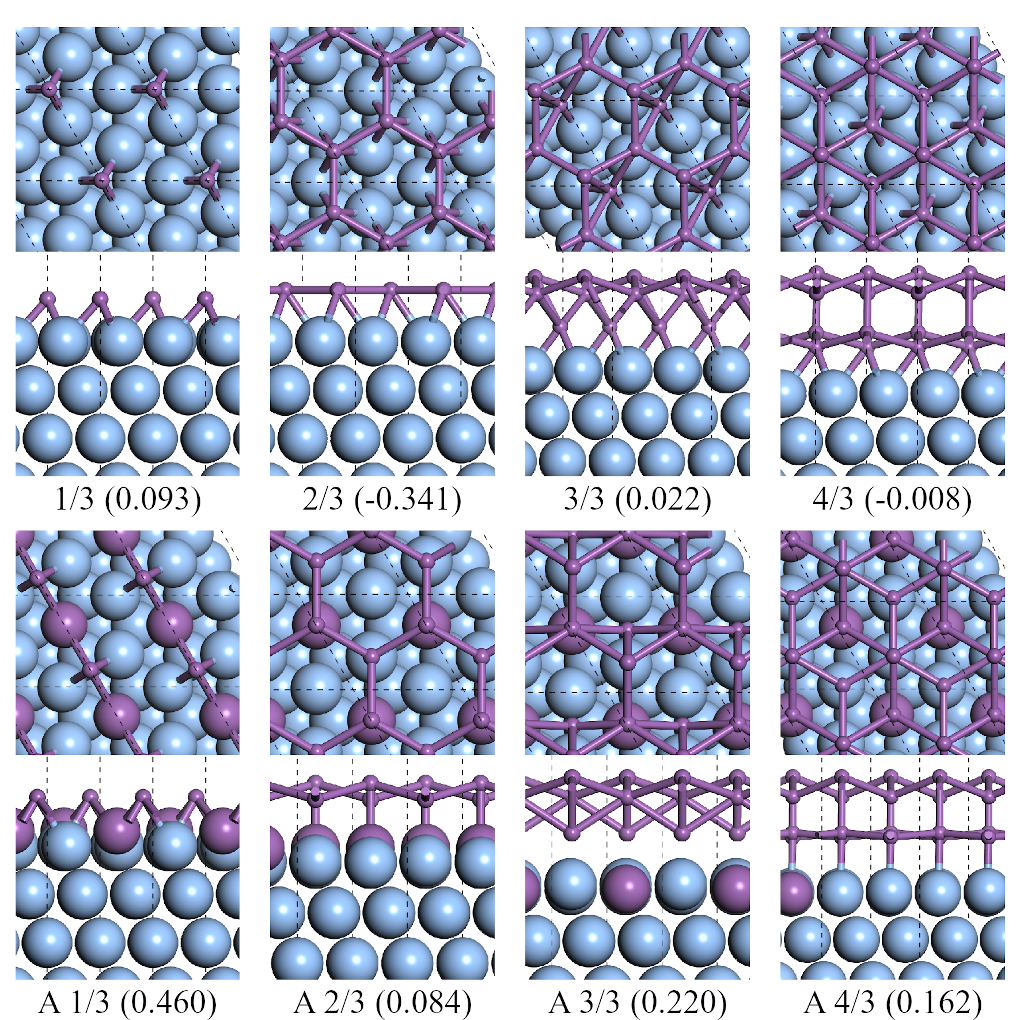}
    \caption{
        Top and side views of the lowest formation energy favorable configurations of antimony adsorbed on $\sqrt{3} \times \sqrt{3}$ Ag(111) and Ag$_2$Sb/Ag(111) substrate. The four sub-figures above correspond to the pure Ag(111) substrate, and the last four sub-figures labeled with `A' correspond to the Ag$_2$Sb/Ag(111) substrate. The fractions under each sub-figure correspond to the coverage of antimony $C_{\text{Sb}}$, and the decimals in brackets represent the formation energies $E_{\text{form}}$. The purple and blue balls represent to the Sb and Ag atoms, respectively.
    }
    \label{fig:struc-r3}
\end{figure*}

The upper four sub-figures in Fig.~\ref{fig:struc-r3} shows the structures with the lowest energy at different coverage of antimony on an $\mathcal{A}$-type pure Ag(111) substrate. Overall, as $C_\text{Sb}$ increased from 1/3 to 4/3, there were two kinds of antimony sheets growing, one with the flat honeycomb lattice at $C_\text{Sb}$ = 2/3, and the other with the buckled honeycomb lattice at $C_\text{Sb}$ = 3/3 and 4/3.
At $C_\text{Sb}$ = 1/3, the Sb atoms sit on the fcc-hollow sites which is the most favorable adsorbed site. According to our tests, the fcc- and hcp-hollow sites have a very similar adsorbed energies, about different of 0.04 eV/atom. Therefore, both fcc- and hcp-hollow sites could be stable during actual production.
At $C_\text{Sb}$ = 2/3, the Sb atoms of the lowest energy configuration sit on the hcp-hollow sites, and formed a flat honeycomb lattice with a 23.1\% stretched lattice constant of $a$ = 5.067 \AA{} compared to freestanding $\beta$-Sb(4.117 \AA{}). % with $d_{\text{Sb-Sb}}$ = 2.924 \AA{}. 
This configuration is the most stable one on $\mathcal{A}$-type Ag(111) substrate and well agreed with the results of the work by Shao et al.~\cite{shaoEpitaxialGrowthFlat2018}.
It is easy to understand that the antimony lattice is flattened by such a large stretch, and it is interesting to note that the $C_{\text{Sb}}$ = 3/3 structure is split into two layers, with the upper layer being a buckled honeycomb lattice and the discrete Sb atoms in the lower layer sitting almost on the bridge sites.
When $C_{\text{Sb}}$ was increased to 4/3, a two-layer buckling honeycomb lattice with AB stacks of different buckling heights was formed. The buckling height of the upper layer is 0.968 \AA{} and the lower layer is 0.712 \AA{}. The electron localization function of C$_{\text{Sb}}$ from 2/3 to 4/3 in Fig.~S1 showed that the electron were localized around Sb atoms, the interaction between Ag-Sb was weaker than Sb-Sb.

% \begin{figure}[htbp]
%     \centering
%     \includegraphics[width=0.99\linewidth]{img/structures-r3.png}
%     \caption{
%         Top and side views of the lowest formation energy favorable configurations of antimony adsorbed on $\sqrt{3} \times \sqrt{3}$ Ag(111) and Ag$_2$Sb/Ag(111) substrate. The four sub-figures above correspond to the pure Ag(111) substrate, and the last four sub-figures labeled with `A' correspond to the Ag$_2$Sb/Ag(111) substrate. The fractions under each sub-figure correspond to the coverage of antimony $C_{\text{Sb}}$, and the decimals in brackets represent the formation energies $E_{\text{form}}$. The purple and blue balls represent to the Sb and Ag atoms, respectively.
%     }
%     \label{fig:struc-r3}
% \end{figure}

As it shown in Fig.~\ref{fig:struc-3r3}, the antimony atoms constructed an $\alpha$-Sb at $C_\text{Sb}$ = 8/6  layer by layer. From C$_\text{Sb}$ = 1/6 to 4/6 on $\mathcal{B}$-type Ag(111) substrate, there existed a clear path forming half layer of $\alpha$-Sb. From 5/6 to 8/6, all configurations were based on $C_{\text{Sb}}$ = 4/6. As Sb continued to be deposited, the atoms were regularly arranged on this half-layer structure, and when $C_{\text{Sb}} = 8/6$, a full layer of $\alpha$-Sb was formed.
The lattice constants of $\alpha$-Sb on Ag(111) substrate are $a$ = 4.388 \AA{} and $b$ = 5.067 \AA{}, which have a stretch of 0.9\% and 6.6\% along each of the $a$(Zigzag) and $b$(Armchair) directions comparing to the freestanding $\alpha$-Sb. The thickness of $\alpha$-Sb was 2.95 \AA{} and the sub-layers of it were almost flat.

% \begin{figure}[htbp]
%     \centering
%     \includegraphics[width=0.99\linewidth]{img/structures-3r3.png}
%     \caption{
%         Top and side views of the lowest formation energy favorable configurations of antimony adsorbed on $3 \times \sqrt{3}$ Ag(111) surface.
%     }
%     \label{fig:struc-3r3}
% \end{figure}

On $\mathcal{C}$-type Ag(111) substrates, we did calculations for Sb coverage from 1/9 to 8/9, a range that contain a monolayer of antimony sheet. Same as $\mathcal{A}$- and $\mathcal{B}$-type Ag(111) substrates, the single Sb sit on the hollow sites. From $C_\text{Sb}$ = 3/9 to 6/9, the most stable structure of each coverage are triangular, quadrilateral, pentagonal and hexagonal lattice, respectively. This is quite different from the results of other two types of substrate. Small flat Sb clusters seem to dominate in the early stages of growth.

% \begin{figure}[htbp]
%     \centering
%     \includegraphics[width=0.99\linewidth]{img/structures-3x3.png}
%     \caption{
%         Top and side views of the lowest formation energy favorable configurations of antimony adsorbed on $3 \times 3$ Ag$_2$Sb/Ag(111) substrate.
%     }
%     \label{fig:struc-3x3}
% \end{figure}

\begin{figure*}[htbp]
    \centering
    \includegraphics[width=\linewidth]{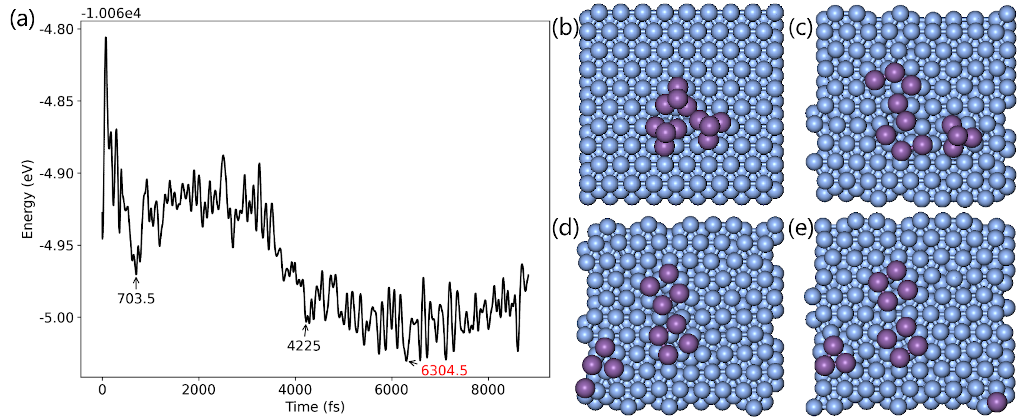}
    \caption{(a) Total energy of system as a function of time from the AIMD simulation for three Sb$_4$ clusters on pure Ag(111) surface at 300K and zero pressure;
        System structures at (b) 0 ps (initial configuration), (c) 703.5ps (two of Sb$_4$ clusters collapse to planar configuration), (d) 4225 ps (all three Sb$_4$ clusters collapse to planar configuration) and (e) 6304.5 ps (the configuration with the lowest energy) are listed at right, respectively.}
    \label{fig:sb12}
\end{figure*}

To further examine the structure of antimony clusters grown on the surface of Ag(111), we calculated the variation of the structure of 1-3 Sb$_4$ clusters falling on the Ag(111) surface by using ab initio molecular dynamics(AIMD) embedded in the package CP2K\cite{kuhneCP2KElectronicStructure2020}.
The substrate of Ag(111) was set to 3 layers with the bottom layer fixed, and the orthogonal lattice parameter was set to $a$ = 26.327 \AA{}, $b$ = 25.333 \AA{}, $c$ = 32.314 \AA{} to avoid the interaction between periodic clusters.
After more than 5 ps at 300 K, the AIMD results showed the process of how small Sb$_4$ clusters spread over the Ag (111) substrate into flat clusters.
Fig.~\ref{fig:sb12}(a) plots the system total energy as a function of time for three Sb$_4$ clusters on pure Ag(111) surface. Near 0.7 ps, two of Sb$_4$ clusters disintegrate and rapidly tile to the substrate,
and at about 4.2 ps, all clusters were spread out. We found that the flat tetragonal clusters exhibit an unusual stability during the simulation. To analysis the interaction between the Sb and Ag atoms, we have drawn the curve of radial distribution function (RDF), see Fig.~S4, it can be seen that the average distance between Sb-Sb stay within 2.935 \AA{}, meanwhile, the Sb-Ag bond increases slightly with the increasing of the number of Sb atoms, from about 2.815 to 2.885 \AA{}. It means that the interaction of Sb clusters with the substrate weakens as the small clusters formed due to the increased inner stability of clusters.
The results of other two configurations were listed in Fig.~S2 and S3 in supplementary information.
We have also examined the integration curve of RDF, the average coordination number of these three system increase from 1 to 2 with the increase of the number of Sb$_4$ clusters.
This can be clearly seen from its structural variations, where dimer, chain and polygonal structures occupy the major conformations.
The flat square Sb$_4$ clusters are very stable in this simulation.
This is consistent with the fact that the most stable configuration on $\mathcal{C}$ type Ag(111) substrate in Fig.~\ref{fig:eform} is Sb$_4$ at $C_{\text{Sb}}$ = 4/9.

% \begin{figure}[htbp]
%     \centering
%     \includegraphics[width=\linewidth]{img/Sb12.png}
%     \caption{(a) Total energy of system as a function of time from the AIMD simulation for three Sb$_4$ clusters on pure Ag(111) surface at 300K and zero pressure;
%         System structures at (b) 0 fs (initial configuration), (c) 703.5 fs (two of Sb$_4$ clusters collapse to planar configuration), (d) 4225 fs (all three Sb$_4$ clusters collapse to planar configuration) and (e) 6304.5 fs (the configuration with the lowest energy) are listed at right, respectively.}
%     \label{fig:sb12}
% \end{figure}

\begin{figure*}[htbp]
    \centering
    \includegraphics[width=0.99\linewidth]{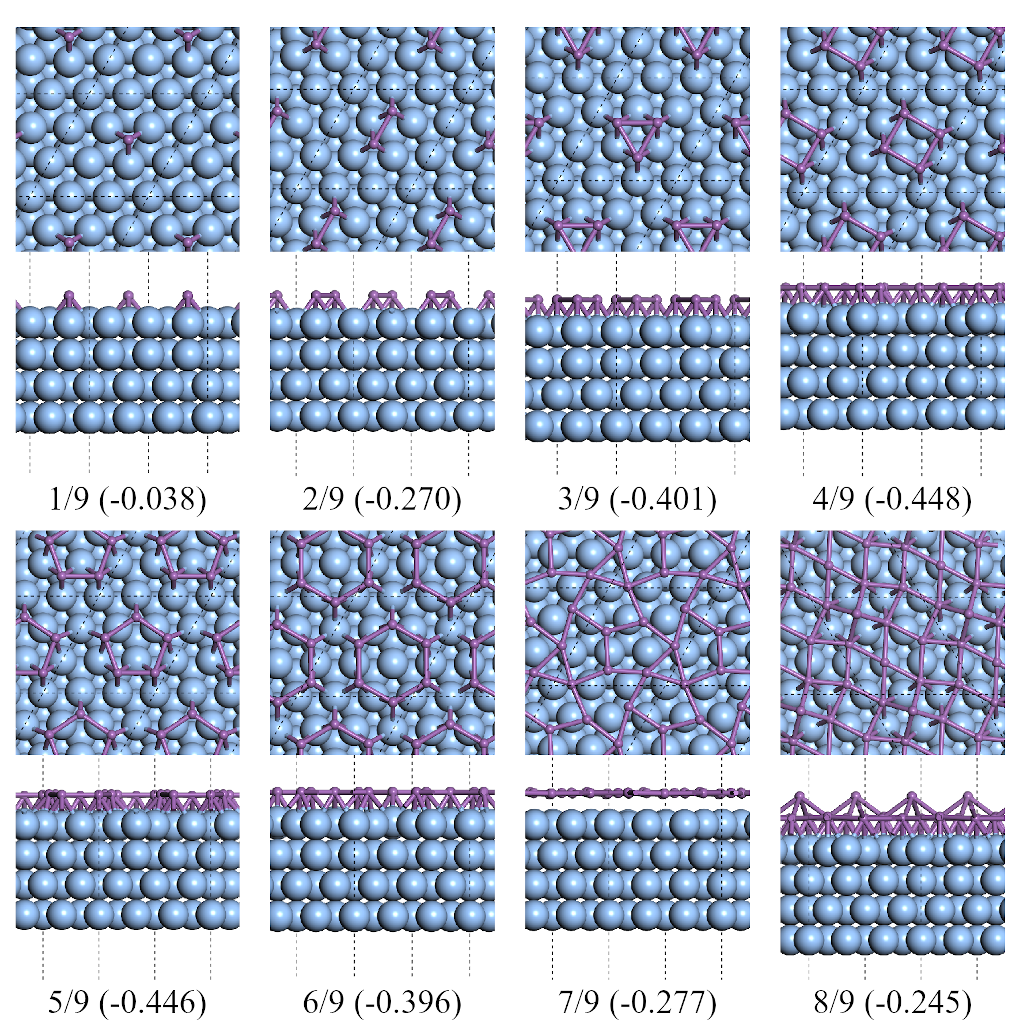}
    \caption{
        Top and side views of the lowest formation energy favorable configurations of antimony adsorbed on $3 \times 3$ Ag$_2$Sb/Ag(111) substrate.
    }
    \label{fig:struc-3x3}
\end{figure*}

The antimony atoms did not spread over the entire surface until $C_{\text{Sb}} = 7/9$.
With the increase of $C_{\text{Sb}}$, the formation energy shows an overall increasing trend, the evolution of the lowest energy structure from different coverages does not allow the formation of a continuous change in the path to $\beta$-Sb, but a valley point appears at $C_{\text{Sb}} = 8/9$, which indicates that the structure is more stable at the coverage. However, the lowest energy structure of $C_{\text{Sb}} = 8/9$ we obtained from structural search is a bilayer structure as shown in the Fig.~\ref{fig:struc-3x3}, with an irregular mesh of 7 atoms in the bottom layer and one atom in the upper layer which located at the center of a square of four atoms of bottom layer. We also obtained the structure of $\beta$-Sb at $C_{\text{Sb}} = 8/9$ with lattice constant of about 4.390 \AA{} and thickness of 1.552 \AA{} (see Fig.~S5, its formation energy is 0.035 eV/atom higher than that of the lowest one. These results indicated that $C_{\text{Sb}} = 8/9$ may be a magic coverage.

Theoretically, the larger the substrate, the closer the result to reality.
Combining the results of these three substrates it can be seen that at lower coverages, the flat 2D clusters occupy the major part, and when the antimony atoms increase enough to spread over the substrate or even more, it is very possible to grow the so-called $\beta$- and $\alpha$-Sb structures. However, in terms of energy comparison, the most stable configuration is the flat half layer of $\alpha$-Sb.

\subsection{Antimonene on Ag(111) with Ag\texorpdfstring{$_2$}{2}Sb surface alloy}\label{sec:res:2}

\begin{figure*}[htbp]
    \centering
    \includegraphics[width=.7\linewidth]{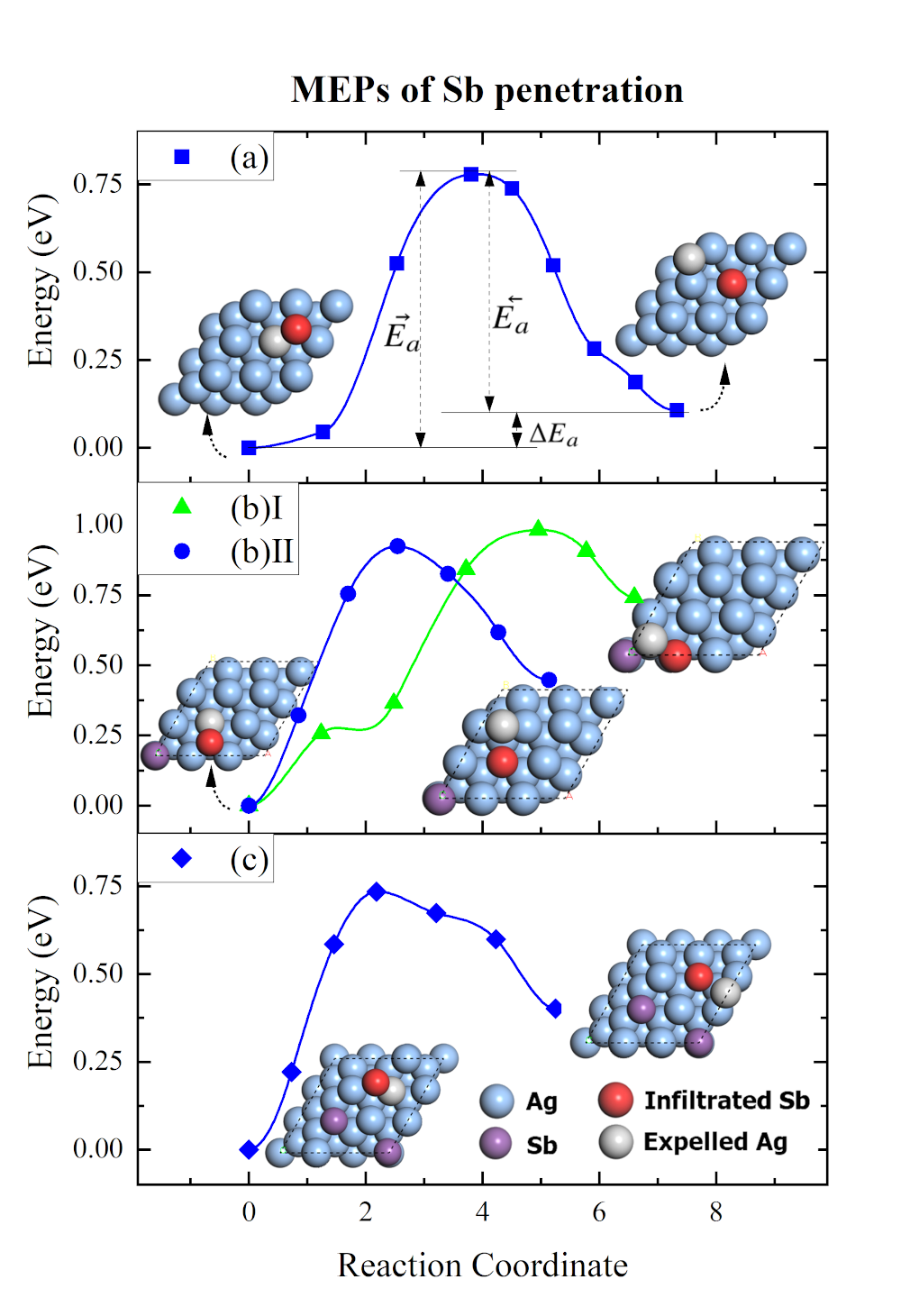}
    \caption{Left and right panels: relaxed structures of (a) one Sb atom penetrate into surface, (b) another Sb atom penetrate into (a) final-state surface, and (c) one more Sb atom penetrate into (b) sub-neighbor site structure. Middle panel plots the MEPs of Sb atoms continuously replace Ag atom on Ag(111) substrate. In these calculations, the expelled Ag atoms were taken off in the next step calculations.} \label{fig:mep}
\end{figure*}

Before calculating the deposition process of Sb on Ag$_2$Sb/Ag(111) substrate, the transition state search method was applied to analyze the process of Sb penetration and replacement of Ag on Ag(111) surface and identify the structure of surface alloy.
We started with modeling the Sb/Ag(111) interface by adding a single Sb atom onto $3\times 3$ supercell Ag(111) surface (see the structure diagram on the left in Fig.~\ref{fig:mep} (a)). The adsorption energy was calculated by the formula:
$E_{\text{ads}}^{\text{Sb}} = E^{\text{Ag(111)}} + E_{\text{atom}}^{\text{Sb}} - E_{\text{tot}}^{\text{Sb/Ag(111)}}, $
where, $E_{\text{atom}}^{\text{Sb}}$ is the total energy of per Sb atom in bulk, and the calculated $E_{\text{ads}}^{\text{Sb}}$ value is 0.037 eV on $3\times 3$ Ag(111) surface. % 4.39
The minimum energy path(MEP) of the penetration of one Sb atom into Ag(111) surface plotted in Fig.~\ref{fig:mep} (a), it shows a forward energy barrier of $\vec{E_a}$ = 0.779 eV, and a reverse energy barrier of $\cev{E_a}$ = 0.672 eV. The $\vec{E_a}$ is much higher than $E_{\text{ads}}^{\text{Sb}}$, therefore, the replacement process should not occur until the environment provides enough energy to allow it to cross the energy barrier, and the generation of surface alloy is fully controllable.

% \begin{figure}[htbp]
%     \centering
%     \includegraphics[width=.99\linewidth]{img/mep.png}
%     \caption{Left and right panels: relaxed structures of (a) one Sb atom penetrate into surface, (b) another Sb atom penetrate into (a) final-state surface, and (c) one more Sb atom penetrate into (b) sub-neighbor site structure. Middle panel plots the MEPs of Sb atoms continuously replace Ag atom on Ag(111) substrate. In these calculations, the expelled Ag atoms were taken off in the next step calculations.} \label{fig:mep}
% \end{figure}

Next, we added a second Sb atom to the substrate while removing the Ag that was expelled. Since there was already a penetrated Sb atom after step (a), so we chose to calculate the MEPs for the penetration at the neighbor and sub-neighbor site separately, see Fig.~\ref{fig:mep} (b). From the results, it is clear that path (b)II is more favorable than path (b)I, with lower energy of the final state and a relatively shorter reaction path. Based on this, we continued to add a 3rd Sb atom to the final state of (b)II and made it penetrate at the sub-neighbor site, the MEP of this process plotted in Fig.~\ref{fig:mep} (c). All energies of MEPs are listed in TABLE. \ref{table:Ea}, the $\Delta E_a$ are all positive, which proves that Sb penetration is a heat absorption process. Such sub-neighbor atomic arrangement characteristics predetermine the impossibility of continuous replacement of Ag by Sb, and with Sb penetration occurring all over the surface, an Ag$_2$Sb alloy monolayer is formed. Till now, we have confirmed that if a surface alloy could be formed, and the alloy should be Ag$_2$Sb.

\begin{table*}
    \centering
    \caption{Forward energy barrier $\vec{E_a}$, reverse energy barrier
        $\cev{E_a}$, and energy difference $\Delta E_a$ of the processes described in
        Fig.~\ref{fig:mep}.}
    \label{table:Ea}
    \begin{ruledtabular}
    \begin{tabular}{lccc}
        % \hline
        % \toprule
        Process                                      & $\vec{E_a}$ & $\cev{E_a}$ & $\Delta E_a$ \\
        \colrule
        % \hline
        (a): 1st Sb penetration                      & 0.779       & 0.672       & +0.107       \\
        (b)I: 2nd Sb penetration: neighbor site      & 0.982       & 0.239       & +0.743       \\
        (b)II: 2nd Sb penetration: sub-neighbor site & 0.924       & 0.447       & +0.447       \\
        (c): 3rd Sb penetration: sub-neighbor site   & 0.735       & 0.333       & +0.402       \\
        % \hline
        % \bottomrule
    \end{tabular}
    \end{ruledtabular}
\end{table*}

Same as pure Ag(111) substrates, three types of Ag$_2$Sb/Ag(111) substrates were modeled and the curve of the formation energies vs. $C_{\text{Sb}}$ were plotted in Fig.~\ref{fig:eform}(b).
Overall, the formation energies on $3\times 3$ substrates were generally lower than on the other two types of substrates. Unlike pure Ag(111) substrates, Sb exhibits different growth characteristics on Ag$_2$Sb monolayer, the small antimony clusters are no longer to be more stable and the formation energy continues to decline until at coverage of 8/9. At the $C_\text{Sb} = 8/9$, the Sb layer shows a $2\times 2$ reconstructed $\beta$-phase structure with minor deformation on $\mathcal{C}$-type Ag$_2$Sb alloy monolayer. One of the Sb atoms in the lower sub-layer is affected by the Sb atom of the substrate and shifts upward, causing the original double sub-layers structure to split into three sub-layers from the side view. The lattice constant is $a$ = 4.388 \AA{}, with about 6.58\% stretching compared to the free state of antimonene (4.117 \AA{}). The layer height is about 4.377 \AA{}. This structure is in good agreement with the results of Sun et al.'s work \cite{sunRealizationBuckledAntimonene2020}.

For $\mathcal{A}$ type of Ag$_2$Sb/Ag(111) substrates, the results of single Sb on Ag$_2$Sb/Ag(111) shows that the Ag-Ag bridge site was most favorable, but a monolayer of $\beta$-Sb with layer height of 0.736 \AA{} formed at $C_{\text{Sb}}=2/3$, and Sb atoms occupy the top sites of both Ag and Sb atoms in substrate, see Fig.~\ref{fig:struc-r3}(A 2/3).
At $C_{\text{Sb}}$ = 3/3, the bilayer structure consisted of a layer of $\beta$-Sb and intercalated atoms, while the intercalated atoms were located at the Ag-Ag b-sites of the substrate.
At $C_{\text{Sb}}$ = 4/3, the bilayer structure consisted of two layers of $\beta$-Sb with AB stacking, the bottom layer was almost flat, and the top layer was buckled honeycomb lattice with buckling higher of 1.022 \AA{}.
Although it shows a very good regular arrangement on $\mathcal{A}$-type Ag$_2$Sb/Ag(111) substrate, the formation energies are higher than that of other two types.
On such substrates, the flat honeycomb antimonene cannot exist stably due to the surface alloy and is replaced by a honeycomb structure with buckling.

% \begin{figure}[htbp]
%     \centering
%     \includegraphics[width=0.99\linewidth]{img/structures-a3r3.png}
%     \caption{
%         Top and side views of the lowest formation energy favorable configurations of antimony adsorbed on $3 \times \sqrt{3}$ Ag$_2$Sb/Ag(111) substrate.
%     }
%     \label{fig:struc-a3r3}
% \end{figure}

\begin{figure*}[htbp]
    \centering
    \includegraphics[width=0.99\linewidth]{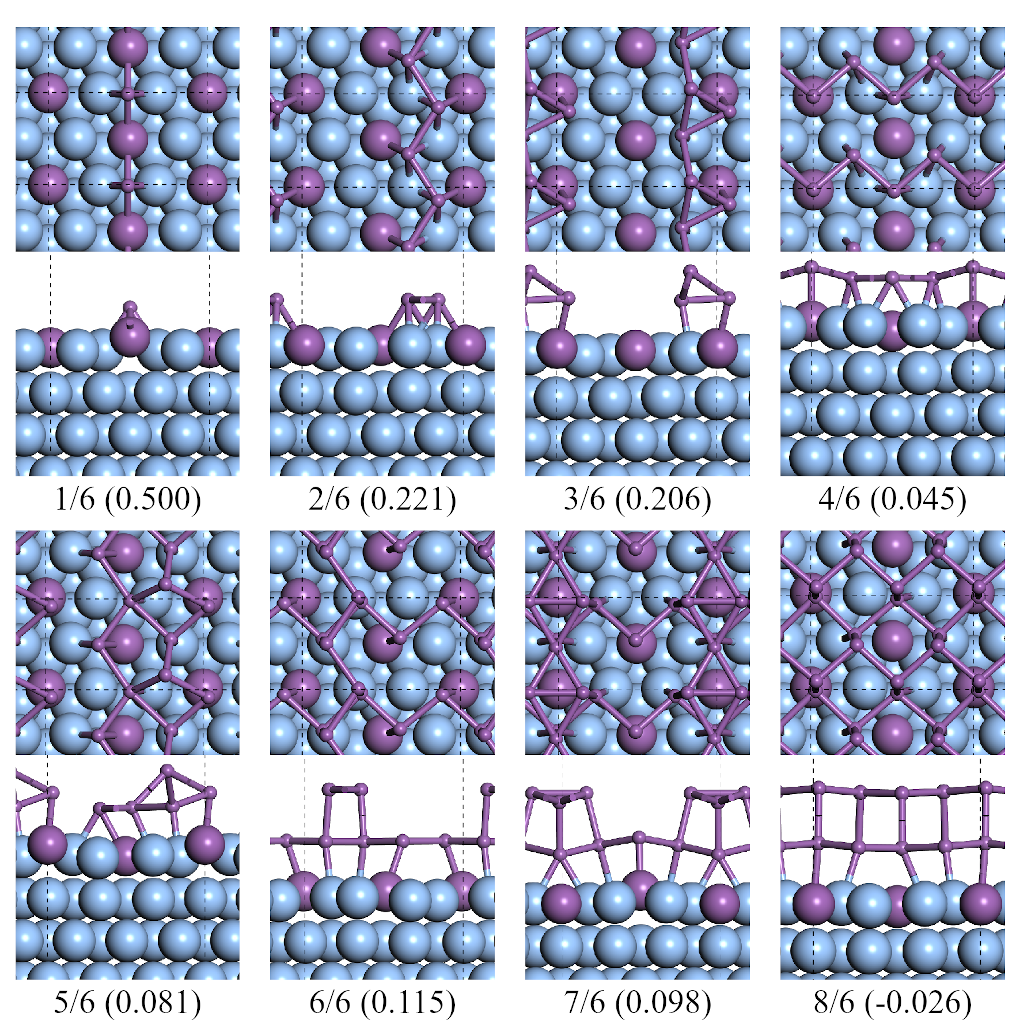}
    \caption{
        Top and side views of the lowest formation energy favorable configurations of antimony adsorbed on $3 \times \sqrt{3}$ Ag$_2$Sb/Ag(111) substrate.
    }
    \label{fig:struc-a3r3}
\end{figure*}

On $\mathcal{B}$-type Ag$_2$Sb/Ag(111) substrates, the lowest formation energy was at $C_{\text{Sb}} = 8/6$, which formed a full layer of $\alpha$-Sb, see Fig.~\ref{fig:struc-a3r3}.
It was almost the same structural evolution path comparing to the pure Ag(111) substrates.
Adsorbed Sb atoms formed a half layer of $\alpha$-Sb first and then growing to full layer.
The difference is that the structure of $\alpha$-Sb was no longer flat, since one of the Sb atoms in the unit cell sit on the top site of the Sb atom in surface alloy while the other Sb atoms sit on the bridge sites of the Ag-Ag pair.

% \begin{figure}[htbp]
%     \centering
%     \includegraphics[width=0.99\linewidth]{img/structures-a3x3.png}
%     \caption{
%         Top and side views of the lowest formation energy favorable configurations of antimony adsorbed on $3 \times 3$ Ag$_2$Sb/Ag(111) substrate.
%     }
%     \label{fig:struc-a3x3}
% \end{figure}

\begin{figure*}[htbp]
    \centering
    \includegraphics[width=0.99\linewidth]{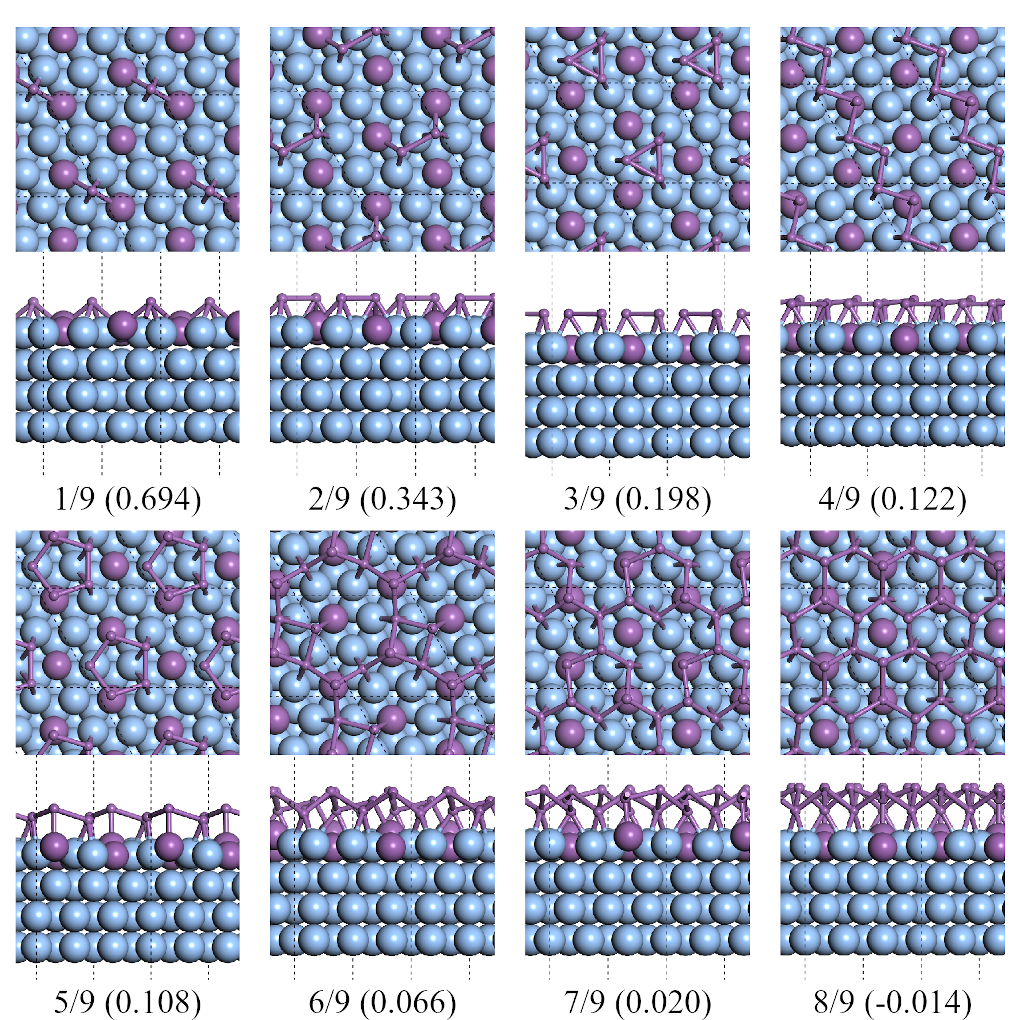}
    \caption{
        Top and side views of the lowest formation energy favorable configurations of antimony adsorbed on $3 \times 3$ Ag$_2$Sb/Ag(111) substrate.
    }
    \label{fig:struc-a3x3}
\end{figure*}

On type $\mathcal{C}$ Ag$_2$Sb/Ag(111) substrates, the lowest formation energy was at $C_{\text{Sb}} = 8/9$ of a reconstructed $2\times 2$ supercell of $\beta$-Sb, see Fig.~\ref{fig:struc-a3x3}. This structure is highly consistent with that in Sun's work \cite{sunRealizationBuckledAntimonene2020}.
From the trend of formation energy, we see that it decreases with the increase of coverage until the most stable structure is formed. The stable small clusters that appear on the pure silver substrate are no longer the most stable structures, although they are also stable structures, due to the influence of the surface alloy.

From the calculation results of the above three substrates we can see that the presence of the surface alloy changes the adsorption site. 
The Sb-Sb interaction is somewhat stronger than that of Sb-Ag, so the presence of Sb atoms in surface alloy can affect the growth of antimonene, especially for the earlier stage. The energetically stable adsorption sites change from the hollow sites to the Ag-Ag bridge sites. Moreover, unlike the pure silver substrate, the Sb top site is also a more stable adsorption site.

\begin{figure*}[htbp]
    \centering
    \includegraphics[width=\linewidth]{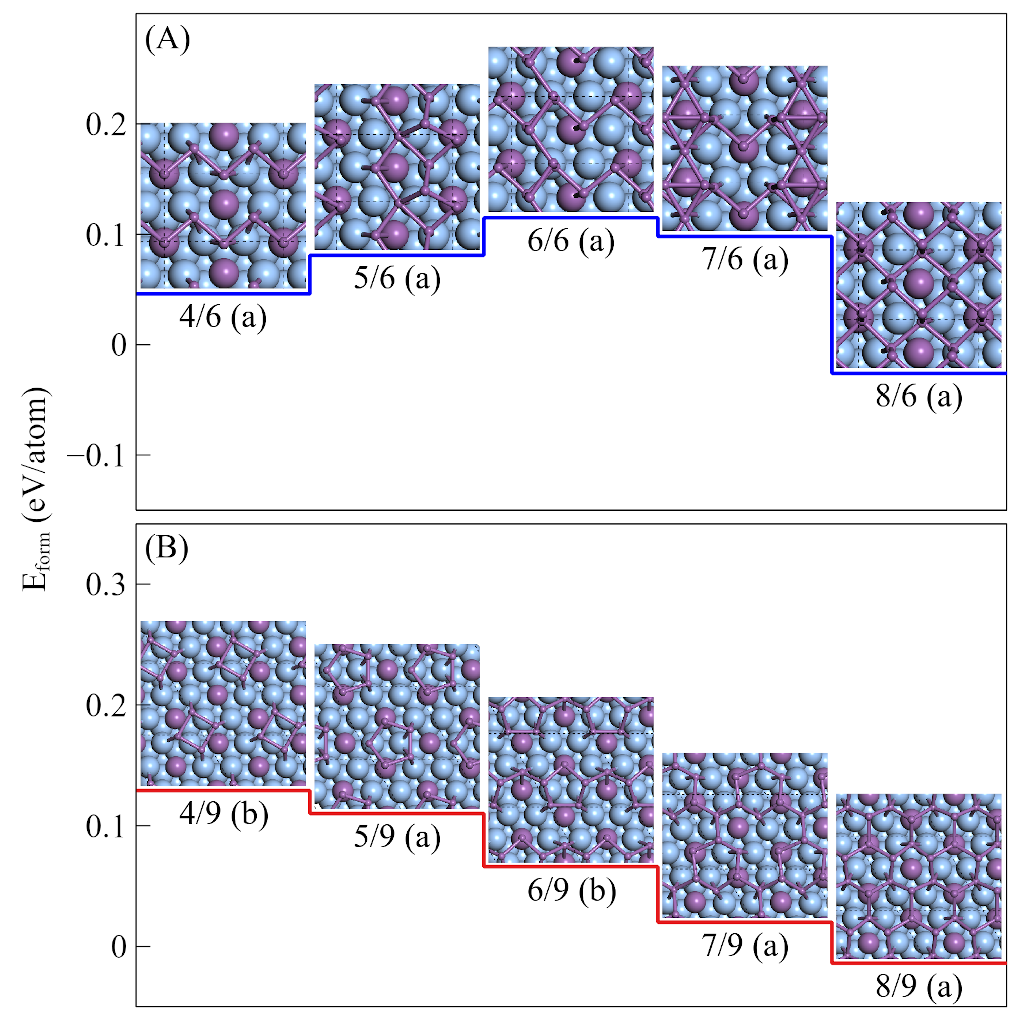}
    \caption{Structural evolution paths of $3\times\sqrt{3}$ and $3\times 3$ supercell surface reconstructions of Sb on Ag$_2$Sb/Ag(111) substrates. (A): $3\times\sqrt{3}$ substrate, culminating in $\alpha$-Sb; (B): $3\times 3$ substrate, culminating in $\beta$-Sb. The fractions marked below each structure represent their coverages, and (a), (b) correspond to the most stable and meta-stable structures, respectively.}
    \label{fig:path}
\end{figure*}

Our calculation shows that the formation energies of $\beta$-Sb (-0.014 eV/atom) and $\alpha$-Sb (-0.026 eV/atom) on Ag$_2$Sb/Ag(111) substrates are very close, so it is difficult to determine who is more stable energetically.
Since both $\alpha$- and $\beta$-Sb monolayers are stable in ambient conditions, We need to compare their structural stability and find their structural growth paths to determine which phrase is easier to growth on and which one is relatively stable.
Therefore, we carefully examined all structures, including the sub-stable ones, from the half-layer to the full-layer and found two evolutionary paths for each phrase on the Ag$_2$Sb/Ag(111) substrate.
In Fig.~\ref{fig:path}(A), the formation energies increased and then decreased with increasing coverage, where the two relatively stable configurations were half- and full-layer $\alpha$-Sb at $C_{\text{Sb}}$ = 4/6 and 8/6, respectively. The structures from $C_\text{Sb}$ = 5/6 to 8/6 all grown based on the stable structure at $C_\text{Sb}$ = 4/6. It is clear that $\alpha$-Sb is grown half-layer by half-layer on the substrate. Although, there exist a small energy barrier of 0.069 eV/atom along this path.
For $3\times 3$ Ag$_2$Sb/Ag(111) substrate in Fig.~\ref{fig:path}(B), at $C_{\text{Sb}}$ = 4/9, 5/9, there were quadrilateral and pentagonal rings with pucker,
in which the Sb$_5$ is a very special one, two of five atoms are sitting on the top site of Sb atoms in substrate, and the other three are on the top site of Ag atoms. This pucker structure is exactly a part of the $\beta$-Sb at C$_{\text{Sb}}$ = 8/9. The continuously deposited Sb atoms placed in their gap positions can form the structures of C$_{\text{Sb}}$ = 6/9 and 7/9, and the structure of C$_{\text{Sb}}$ = 7/9 is exactly $\beta$-Sb with a Sb vacancy.

% \begin{figure}[htbp]
%     \centering
%     \includegraphics[width=\linewidth]{img/StructurePath.png}
%     \caption{Structural evolution paths of $3\times\sqrt{3}$ and $3\times 3$ supercell surface reconstructions of Sb on Ag$_2$Sb/Ag(111) substrates. (A): $3\times\sqrt{3}$ substrate, culminating in $\alpha$-Sb; (B): $3\times 3$ substrate, culminating in $\beta$-Sb. The fractions marked below each structure represent their coverages, and (a), (b) correspond to the most stable and meta-stable structures, respectively.}
%     \label{fig:path}
% \end{figure}

Considering the results of all types of substrates together, both the size of reconstructed supercells  and the surface alloy have a significant effect on growth.
No matter with or without surface alloys, the structure of antimonene on the $\mathcal{A}$-type of substrates always show a large stretch whether forming the flat honeycomb antimonene or low-buckled honeycomb antimonene. On $\mathcal{B}$-type of substrates, the $\alpha$-Sb occupy the main structures, and there exist a very clear evolutionary path of growing of half-layer by half-layer. The half layer $\alpha$-Sb has a high stability, regardless of the deformation by the substrates. 
On the largest substrates in this work, the $\beta$-Sb exhibits excellent stability on substrate with surface alloy and much less stability on pure Ag(111) substrates compared to flat honeycomb and half-layer $\alpha$-phase antimonene. It can be seen that the phase of antimonene is deeply influenced by the size of the reconstructed supercells. 
Especially, at the coverage of 2/3, which is included in the all types of substrates, all configurations are at lower energy levels and the structures of antimonene are significantly modulated by the size of reconstructed supercells. For smaller substrates ($\mathcal{A}$ and $\mathcal{B}$), Sb atoms could fully cover the substrates, but not for larger substrates ($\mathcal{C}$). This is the direct factor that determines the different phases produced by the different size of the reconstructed supercells. If we can control the size of the reconstructed supercells, the phase of antimonene will be determined.

The existence of surface alloy of Ag$_2$Sb has also played an important role in the growth of antimonene. At lower coverages (C$_\text{Sb} < 2/3$), the clusters of Sb$_N$ (N = 3, 4, 5) although present, are no longer the most stable structures. On the other hand, at higher coverages (C$_\text{Sb} > 2/3$), the formation energies of $\alpha$- and $\beta$-Sb shift from higher to the lowest level with the present of surface alloy. This has a positive effect on the formation of stable monolayer of antimonene.

Note that the minimum value of required coverage to form a monolayer are 2/3, 2/3, and 7/9 for substrate of $\mathcal{A}$, $\mathcal{B}$ and $\mathcal{C}$ types, respectively. Our simulations are based on the complete relaxation of the Sb atoms on the substrate, and if the speed of the process of atomic relaxation can be controlled, then the phase of antimonene should also be controllable. On molecular beam epitaxy experiments, the control of relaxation is usually multifactorial, such as temperature, pressure, source evaporation rate, etc. Which one is the main factor affecting the relaxation speed remains to be finely verified.

\section{Conclusion}\label{conclusion}
In summary, we have simulated the growth process of antimonene on Ag(111) substrate with and without Ag$_2$Sb surface alloy using DFT and PSO methods. According to the results, in addition to the experimental reported flat honeycomb and $\beta$-phase antimonene, $\alpha$-phase antimonene had also been observed to growth on thus substrates and showed a highly stable topological structure with low puckering. Comparing the silver substrates with and without surface alloys, it was found that the presence of surface alloys does not hinder the growth of antimonene, but rather promotes the stabilization of antimonene. On substrates with Ag$_2$Sb surface alloy, both $\alpha$-Sb and $\beta$-Sb have been observed to growth. The core of the phase modulation engineering lies in the control of the relaxation rate of antimony deposited atoms.
Our work provides a more comprehensive theoretical insight into the growth of antimonene on silver substrates, complementing possible omissions in experiments that may provide some basis for antimonene synthesis.

\begin{acknowledgments}
    This work is supported by the National Natural Science Foundation of China (Grant Nos. 12174084, 11904075, 11874139), the National Postdoctoral Program for Innovative Talents (Grant No. BX20190104), the Scientific and Technological Research Foundation of Hebei Province (Grant No. ZD2021065), and the Key Program of Natural Science Foundation of Hebei Province (Grant No. A2021205024).
\end{acknowledgments}

\bibliographystyle{apsrev4-2}
\bibliography{refs} % Produces the bibliography via BibTeX.

%apsrev4-2.bst 2019-01-14 (MD) hand-edited version of apsrev4-1.bst
%Control: key (0)
%Control: author (72) initials jnrlst
%Control: editor formatted (1) identically to author
%Control: production of article title (-1) disabled
%Control: page (0) single
%Control: year (1) truncated
%Control: production of eprint (0) enabled
\begin{thebibliography}{44}%
\makeatletter
\providecommand \@ifxundefined [1]{%
 \@ifx{#1\undefined}
}%
\providecommand \@ifnum [1]{%
 \ifnum #1\expandafter \@firstoftwo
 \else \expandafter \@secondoftwo
 \fi
}%
\providecommand \@ifx [1]{%
 \ifx #1\expandafter \@firstoftwo
 \else \expandafter \@secondoftwo
 \fi
}%
\providecommand \natexlab [1]{#1}%
\providecommand \enquote  [1]{``#1''}%
\providecommand \bibnamefont  [1]{#1}%
\providecommand \bibfnamefont [1]{#1}%
\providecommand \citenamefont [1]{#1}%
\providecommand \href@noop [0]{\@secondoftwo}%
\providecommand \href [0]{\begingroup \@sanitize@url \@href}%
\providecommand \@href[1]{\@@startlink{#1}\@@href}%
\providecommand \@@href[1]{\endgroup#1\@@endlink}%
\providecommand \@sanitize@url [0]{\catcode `\\12\catcode `\$12\catcode
  `\&12\catcode `\#12\catcode `\^12\catcode `\_12\catcode `\%12\relax}%
\providecommand \@@startlink[1]{}%
\providecommand \@@endlink[0]{}%
\providecommand \url  [0]{\begingroup\@sanitize@url \@url }%
\providecommand \@url [1]{\endgroup\@href {#1}{\urlprefix }}%
\providecommand \urlprefix  [0]{URL }%
\providecommand \Eprint [0]{\href }%
\providecommand \doibase [0]{https://doi.org/}%
\providecommand \selectlanguage [0]{\@gobble}%
\providecommand \bibinfo  [0]{\@secondoftwo}%
\providecommand \bibfield  [0]{\@secondoftwo}%
\providecommand \translation [1]{[#1]}%
\providecommand \BibitemOpen [0]{}%
\providecommand \bibitemStop [0]{}%
\providecommand \bibitemNoStop [0]{.\EOS\space}%
\providecommand \EOS [0]{\spacefactor3000\relax}%
\providecommand \BibitemShut  [1]{\csname bibitem#1\endcsname}%
\let\auto@bib@innerbib\@empty
%</preamble>
\bibitem [{\citenamefont {Ares}\ \emph {et~al.}(2018)\citenamefont {Ares},
  \citenamefont {Palacios}, \citenamefont {Abell{\'a}n}, \citenamefont
  {{G{\'o}mez-Herrero}},\ and\ \citenamefont
  {Zamora}}]{aresRecentProgressAntimonene2018}%
  \BibitemOpen
  \bibfield  {author} {\bibinfo {author} {\bibfnamefont {P.}~\bibnamefont
  {Ares}}, \bibinfo {author} {\bibfnamefont {J.~J.}\ \bibnamefont {Palacios}},
  \bibinfo {author} {\bibfnamefont {G.}~\bibnamefont {Abell{\'a}n}}, \bibinfo
  {author} {\bibfnamefont {J.}~\bibnamefont {{G{\'o}mez-Herrero}}},\ and\
  \bibinfo {author} {\bibfnamefont {F.}~\bibnamefont {Zamora}},\ }\href
  {https://doi.org/10.1002/adma.201703771} {\bibfield  {journal} {\bibinfo
  {journal} {Advanced Materials}\ }\textbf {\bibinfo {volume} {30}},\ \bibinfo
  {pages} {1703771} (\bibinfo {year} {2018})}\BibitemShut {NoStop}%
\bibitem [{\citenamefont {Zhang}\ \emph {et~al.}(2018)\citenamefont {Zhang},
  \citenamefont {Guo}, \citenamefont {Chen}, \citenamefont {Wang},
  \citenamefont {Gao}, \citenamefont {{G{\'o}mez-Herrero}}, \citenamefont
  {Ares}, \citenamefont {Zamora}, \citenamefont {Zhu},\ and\ \citenamefont
  {Zeng}}]{zhangRecentProgress2D2018}%
  \BibitemOpen
  \bibfield  {author} {\bibinfo {author} {\bibfnamefont {S.}~\bibnamefont
  {Zhang}}, \bibinfo {author} {\bibfnamefont {S.}~\bibnamefont {Guo}}, \bibinfo
  {author} {\bibfnamefont {Z.}~\bibnamefont {Chen}}, \bibinfo {author}
  {\bibfnamefont {Y.}~\bibnamefont {Wang}}, \bibinfo {author} {\bibfnamefont
  {H.}~\bibnamefont {Gao}}, \bibinfo {author} {\bibfnamefont {J.}~\bibnamefont
  {{G{\'o}mez-Herrero}}}, \bibinfo {author} {\bibfnamefont {P.}~\bibnamefont
  {Ares}}, \bibinfo {author} {\bibfnamefont {F.}~\bibnamefont {Zamora}},
  \bibinfo {author} {\bibfnamefont {Z.}~\bibnamefont {Zhu}},\ and\ \bibinfo
  {author} {\bibfnamefont {H.}~\bibnamefont {Zeng}},\ }\href
  {https://doi.org/10.1039/c7cs00125h} {\bibfield  {journal} {\bibinfo
  {journal} {Chemical Society Reviews}\ }\textbf {\bibinfo {volume} {47}},\
  \bibinfo {pages} {982} (\bibinfo {year} {2018})}\BibitemShut {NoStop}%
\bibitem [{\citenamefont {Wang}\ \emph {et~al.}(2019)\citenamefont {Wang},
  \citenamefont {Song},\ and\ \citenamefont
  {Qu}}]{wangAntimoneneExperimentalPreparation2019}%
  \BibitemOpen
  \bibfield  {author} {\bibinfo {author} {\bibfnamefont {X.}~\bibnamefont
  {Wang}}, \bibinfo {author} {\bibfnamefont {J.}~\bibnamefont {Song}},\ and\
  \bibinfo {author} {\bibfnamefont {J.}~\bibnamefont {Qu}},\ }\href
  {https://doi.org/10.1002/anie.201808302} {\bibfield  {journal} {\bibinfo
  {journal} {Angewandte Chemie International Edition}\ }\textbf {\bibinfo
  {volume} {58}},\ \bibinfo {pages} {1574} (\bibinfo {year}
  {2019})}\BibitemShut {NoStop}%
\bibitem [{\citenamefont {Zhang}\ \emph {et~al.}(2015)\citenamefont {Zhang},
  \citenamefont {Yan}, \citenamefont {Li}, \citenamefont {Chen},\ and\
  \citenamefont {Zeng}}]{zhangAtomicallyThinArsenene2015}%
  \BibitemOpen
  \bibfield  {author} {\bibinfo {author} {\bibfnamefont {S.}~\bibnamefont
  {Zhang}}, \bibinfo {author} {\bibfnamefont {Z.}~\bibnamefont {Yan}}, \bibinfo
  {author} {\bibfnamefont {Y.}~\bibnamefont {Li}}, \bibinfo {author}
  {\bibfnamefont {Z.}~\bibnamefont {Chen}},\ and\ \bibinfo {author}
  {\bibfnamefont {H.}~\bibnamefont {Zeng}},\ }\href
  {https://doi.org/10.1002/anie.201411246} {\bibfield  {journal} {\bibinfo
  {journal} {Angewandte Chemie International Edition}\ }\textbf {\bibinfo
  {volume} {54}},\ \bibinfo {pages} {3112} (\bibinfo {year}
  {2015})}\BibitemShut {NoStop}%
\bibitem [{\citenamefont {Wang}\ \emph {et~al.}(2015)\citenamefont {Wang},
  \citenamefont {Pandey},\ and\ \citenamefont
  {Karna}}]{wangAtomicallyThinGroup2015}%
  \BibitemOpen
  \bibfield  {author} {\bibinfo {author} {\bibfnamefont {G.}~\bibnamefont
  {Wang}}, \bibinfo {author} {\bibfnamefont {R.}~\bibnamefont {Pandey}},\ and\
  \bibinfo {author} {\bibfnamefont {S.~P.}\ \bibnamefont {Karna}},\ }\href
  {https://doi.org/10.1021/acsami.5b02441} {\bibfield  {journal} {\bibinfo
  {journal} {ACS Applied Materials \& Interfaces}\ }\textbf {\bibinfo {volume}
  {7}},\ \bibinfo {pages} {11490} (\bibinfo {year} {2015})}\BibitemShut
  {NoStop}%
\bibitem [{\citenamefont {Wang}\ and\ \citenamefont
  {Ding}(2015)}]{wangElectronicStructureCarrier2015}%
  \BibitemOpen
  \bibfield  {author} {\bibinfo {author} {\bibfnamefont {Y.}~\bibnamefont
  {Wang}}\ and\ \bibinfo {author} {\bibfnamefont {Y.}~\bibnamefont {Ding}},\
  }\href {https://doi.org/10.1186/s11671-015-0955-7} {\bibfield  {journal}
  {\bibinfo  {journal} {Nanoscale Research Letters}\ }\textbf {\bibinfo
  {volume} {10}},\ \bibinfo {pages} {254} (\bibinfo {year} {2015})}\BibitemShut
  {NoStop}%
\bibitem [{\citenamefont {Akt{\"u}rk}\ \emph {et~al.}(2015)\citenamefont
  {Akt{\"u}rk}, \citenamefont {{\"O}z{\c c}elik},\ and\ \citenamefont
  {Ciraci}}]{akturkSinglelayerCrystallinePhases2015}%
  \BibitemOpen
  \bibfield  {author} {\bibinfo {author} {\bibfnamefont {O.~{\"U}.}\
  \bibnamefont {Akt{\"u}rk}}, \bibinfo {author} {\bibfnamefont {V.~O.}\
  \bibnamefont {{\"O}z{\c c}elik}},\ and\ \bibinfo {author} {\bibfnamefont
  {S.}~\bibnamefont {Ciraci}},\ }\href
  {https://doi.org/10.1103/physrevb.91.235446} {\bibfield  {journal} {\bibinfo
  {journal} {Physical Review B}\ }\textbf {\bibinfo {volume} {91}},\ \bibinfo
  {pages} {235446} (\bibinfo {year} {2015})}\BibitemShut {NoStop}%
\bibitem [{\citenamefont {Wang}\ \emph {et~al.}(2016)\citenamefont {Wang},
  \citenamefont {Wang},\ and\ \citenamefont
  {Zhao}}]{wangThermalTransportProperties2016}%
  \BibitemOpen
  \bibfield  {author} {\bibinfo {author} {\bibfnamefont {S.}~\bibnamefont
  {Wang}}, \bibinfo {author} {\bibfnamefont {W.}~\bibnamefont {Wang}},\ and\
  \bibinfo {author} {\bibfnamefont {G.}~\bibnamefont {Zhao}},\ }\href
  {https://doi.org/10.1039/c6cp06088a} {\bibfield  {journal} {\bibinfo
  {journal} {Physical Chemistry Chemical Physics}\ }\textbf {\bibinfo {volume}
  {18}},\ \bibinfo {pages} {31217} (\bibinfo {year} {2016})}\BibitemShut
  {NoStop}%
\bibitem [{\citenamefont {Zhao}\ \emph {et~al.}(2015)\citenamefont {Zhao},
  \citenamefont {Zhang},\ and\ \citenamefont
  {Li}}]{zhaoStraindrivenBandInversion2015}%
  \BibitemOpen
  \bibfield  {author} {\bibinfo {author} {\bibfnamefont {M.}~\bibnamefont
  {Zhao}}, \bibinfo {author} {\bibfnamefont {X.}~\bibnamefont {Zhang}},\ and\
  \bibinfo {author} {\bibfnamefont {L.}~\bibnamefont {Li}},\ }\href
  {https://doi.org/10.1038/srep16108} {\bibfield  {journal} {\bibinfo
  {journal} {Scientific Reports}\ }\textbf {\bibinfo {volume} {5}},\ \bibinfo
  {pages} {16108} (\bibinfo {year} {2015})}\BibitemShut {NoStop}%
\bibitem [{\citenamefont {Kripalani}\ \emph {et~al.}(2018)\citenamefont
  {Kripalani}, \citenamefont {Kistanov}, \citenamefont {Cai}, \citenamefont
  {Xue},\ and\ \citenamefont
  {Zhou}}]{kripalaniStrainEngineeringAntimonene2018}%
  \BibitemOpen
  \bibfield  {author} {\bibinfo {author} {\bibfnamefont {D.~R.}\ \bibnamefont
  {Kripalani}}, \bibinfo {author} {\bibfnamefont {A.~A.}\ \bibnamefont
  {Kistanov}}, \bibinfo {author} {\bibfnamefont {Y.}~\bibnamefont {Cai}},
  \bibinfo {author} {\bibfnamefont {M.}~\bibnamefont {Xue}},\ and\ \bibinfo
  {author} {\bibfnamefont {K.}~\bibnamefont {Zhou}},\ }\href
  {https://doi.org/10.1103/physrevb.98.085410} {\bibfield  {journal} {\bibinfo
  {journal} {Physical Review B}\ }\textbf {\bibinfo {volume} {98}},\ \bibinfo
  {pages} {085410} (\bibinfo {year} {2018})}\BibitemShut {NoStop}%
\bibitem [{\citenamefont {Zhao}\ and\ \citenamefont
  {Wang}(2020)}]{zhaoTwodimensionalGraphenelikeXenes2020}%
  \BibitemOpen
  \bibfield  {author} {\bibinfo {author} {\bibfnamefont {A.}~\bibnamefont
  {Zhao}}\ and\ \bibinfo {author} {\bibfnamefont {B.}~\bibnamefont {Wang}},\
  }\href {https://doi.org/10.1063/1.5135984} {\bibfield  {journal} {\bibinfo
  {journal} {APL Materials}\ }\textbf {\bibinfo {volume} {8}},\ \bibinfo
  {pages} {030701} (\bibinfo {year} {2020})}\BibitemShut {NoStop}%
\bibitem [{\citenamefont {Cheung}\ \emph {et~al.}(2016)\citenamefont {Cheung},
  \citenamefont {Fuh}, \citenamefont {Hsu}, \citenamefont {Lin},\ and\
  \citenamefont {Chang}}]{cheungSpinOrbitCoupling2016}%
  \BibitemOpen
  \bibfield  {author} {\bibinfo {author} {\bibfnamefont {C.-H.}\ \bibnamefont
  {Cheung}}, \bibinfo {author} {\bibfnamefont {H.-R.}\ \bibnamefont {Fuh}},
  \bibinfo {author} {\bibfnamefont {M.-C.}\ \bibnamefont {Hsu}}, \bibinfo
  {author} {\bibfnamefont {Y.-C.}\ \bibnamefont {Lin}},\ and\ \bibinfo {author}
  {\bibfnamefont {C.-R.}\ \bibnamefont {Chang}},\ }\href
  {https://doi.org/10.1186/s11671-016-1666-4} {\bibfield  {journal} {\bibinfo
  {journal} {Nanoscale Research Letters}\ }\textbf {\bibinfo {volume} {11}},\
  \bibinfo {pages} {459} (\bibinfo {year} {2016})}\BibitemShut {NoStop}%
\bibitem [{\citenamefont {Novoselov}(2004)}]{novoselovElectricFieldEffect2004}%
  \BibitemOpen
  \bibfield  {author} {\bibinfo {author} {\bibfnamefont {K.~S.}\ \bibnamefont
  {Novoselov}},\ }\href {https://doi.org/10.1126/science.1102896} {\bibfield
  {journal} {\bibinfo  {journal} {Science}\ }\textbf {\bibinfo {volume}
  {306}},\ \bibinfo {pages} {666} (\bibinfo {year} {2004})}\BibitemShut
  {NoStop}%
\bibitem [{\citenamefont {Ares}\ \emph {et~al.}(2016)\citenamefont {Ares},
  \citenamefont {{Aguilar-Galindo}}, \citenamefont
  {{Rodr{\'i}guez-San-Miguel}}, \citenamefont {Aldave}, \citenamefont
  {{D{\'i}az-Tendero}}, \citenamefont {Alcam{\'i}}, \citenamefont {Mart{\'i}n},
  \citenamefont {{G{\'o}mez-Herrero}},\ and\ \citenamefont
  {Zamora}}]{aresMechanicalIsolationHighly2016}%
  \BibitemOpen
  \bibfield  {author} {\bibinfo {author} {\bibfnamefont {P.}~\bibnamefont
  {Ares}}, \bibinfo {author} {\bibfnamefont {F.}~\bibnamefont
  {{Aguilar-Galindo}}}, \bibinfo {author} {\bibfnamefont {D.}~\bibnamefont
  {{Rodr{\'i}guez-San-Miguel}}}, \bibinfo {author} {\bibfnamefont {D.~A.}\
  \bibnamefont {Aldave}}, \bibinfo {author} {\bibfnamefont {S.}~\bibnamefont
  {{D{\'i}az-Tendero}}}, \bibinfo {author} {\bibfnamefont {M.}~\bibnamefont
  {Alcam{\'i}}}, \bibinfo {author} {\bibfnamefont {F.}~\bibnamefont
  {Mart{\'i}n}}, \bibinfo {author} {\bibfnamefont {J.}~\bibnamefont
  {{G{\'o}mez-Herrero}}},\ and\ \bibinfo {author} {\bibfnamefont
  {F.}~\bibnamefont {Zamora}},\ }\href {https://doi.org/10.1002/adma.201602128}
  {\bibfield  {journal} {\bibinfo  {journal} {Advanced Materials}\ }\textbf
  {\bibinfo {volume} {28}},\ \bibinfo {pages} {6332} (\bibinfo {year}
  {2016})}\BibitemShut {NoStop}%
\bibitem [{\citenamefont {Gibaja}\ \emph {et~al.}(2016)\citenamefont {Gibaja},
  \citenamefont {{Rodriguez-San-Miguel}}, \citenamefont {Ares}, \citenamefont
  {{G{\'o}mez-Herrero}}, \citenamefont {Varela}, \citenamefont {Gillen},
  \citenamefont {Maultzsch}, \citenamefont {Hauke}, \citenamefont {Hirsch},
  \citenamefont {Abell{\'a}n},\ and\ \citenamefont
  {Zamora}}]{gibajaFewLayerAntimoneneLiquidPhase2016}%
  \BibitemOpen
  \bibfield  {author} {\bibinfo {author} {\bibfnamefont {C.}~\bibnamefont
  {Gibaja}}, \bibinfo {author} {\bibfnamefont {D.}~\bibnamefont
  {{Rodriguez-San-Miguel}}}, \bibinfo {author} {\bibfnamefont {P.}~\bibnamefont
  {Ares}}, \bibinfo {author} {\bibfnamefont {J.}~\bibnamefont
  {{G{\'o}mez-Herrero}}}, \bibinfo {author} {\bibfnamefont {M.}~\bibnamefont
  {Varela}}, \bibinfo {author} {\bibfnamefont {R.}~\bibnamefont {Gillen}},
  \bibinfo {author} {\bibfnamefont {J.}~\bibnamefont {Maultzsch}}, \bibinfo
  {author} {\bibfnamefont {F.}~\bibnamefont {Hauke}}, \bibinfo {author}
  {\bibfnamefont {A.}~\bibnamefont {Hirsch}}, \bibinfo {author} {\bibfnamefont
  {G.}~\bibnamefont {Abell{\'a}n}},\ and\ \bibinfo {author} {\bibfnamefont
  {F.}~\bibnamefont {Zamora}},\ }\href {https://doi.org/10.1002/anie.201605298}
  {\bibfield  {journal} {\bibinfo  {journal} {Angewandte Chemie International
  Edition}\ }\textbf {\bibinfo {volume} {55}},\ \bibinfo {pages} {14345}
  (\bibinfo {year} {2016})}\BibitemShut {NoStop}%
\bibitem [{\citenamefont {Wang}\ \emph {et~al.}(2018)\citenamefont {Wang},
  \citenamefont {He}, \citenamefont {Zhou}, \citenamefont {Zhang},
  \citenamefont {Wu}, \citenamefont {Hu}, \citenamefont {Liu}, \citenamefont
  {Song},\ and\ \citenamefont {Qu}}]{wangBandgapTunablePreparationSmooth2018}%
  \BibitemOpen
  \bibfield  {author} {\bibinfo {author} {\bibfnamefont {X.}~\bibnamefont
  {Wang}}, \bibinfo {author} {\bibfnamefont {J.}~\bibnamefont {He}}, \bibinfo
  {author} {\bibfnamefont {B.}~\bibnamefont {Zhou}}, \bibinfo {author}
  {\bibfnamefont {Y.}~\bibnamefont {Zhang}}, \bibinfo {author} {\bibfnamefont
  {J.}~\bibnamefont {Wu}}, \bibinfo {author} {\bibfnamefont {R.}~\bibnamefont
  {Hu}}, \bibinfo {author} {\bibfnamefont {L.}~\bibnamefont {Liu}}, \bibinfo
  {author} {\bibfnamefont {J.}~\bibnamefont {Song}},\ and\ \bibinfo {author}
  {\bibfnamefont {J.}~\bibnamefont {Qu}},\ }\href
  {https://doi.org/10.1002/ange.201804886} {\bibfield  {journal} {\bibinfo
  {journal} {Angewandte Chemie}\ }\textbf {\bibinfo {volume} {130}},\ \bibinfo
  {pages} {8804} (\bibinfo {year} {2018})}\BibitemShut {NoStop}%
\bibitem [{\citenamefont {Gibaja}\ \emph {et~al.}(2019)\citenamefont {Gibaja},
  \citenamefont {Assebban}, \citenamefont {Torres}, \citenamefont {Fickert},
  \citenamefont {{Sanchis-Gual}}, \citenamefont {Brotons}, \citenamefont {Paz},
  \citenamefont {Palacios}, \citenamefont {Michel}, \citenamefont
  {Abell{\'a}n},\ and\ \citenamefont
  {Zamora}}]{gibajaLiquidPhaseExfoliation2019}%
  \BibitemOpen
  \bibfield  {author} {\bibinfo {author} {\bibfnamefont {C.}~\bibnamefont
  {Gibaja}}, \bibinfo {author} {\bibfnamefont {M.}~\bibnamefont {Assebban}},
  \bibinfo {author} {\bibfnamefont {I.}~\bibnamefont {Torres}}, \bibinfo
  {author} {\bibfnamefont {M.}~\bibnamefont {Fickert}}, \bibinfo {author}
  {\bibfnamefont {R.}~\bibnamefont {{Sanchis-Gual}}}, \bibinfo {author}
  {\bibfnamefont {I.}~\bibnamefont {Brotons}}, \bibinfo {author} {\bibfnamefont
  {W.~S.}\ \bibnamefont {Paz}}, \bibinfo {author} {\bibfnamefont {J.~J.}\
  \bibnamefont {Palacios}}, \bibinfo {author} {\bibfnamefont {E.~G.}\
  \bibnamefont {Michel}}, \bibinfo {author} {\bibfnamefont {G.}~\bibnamefont
  {Abell{\'a}n}},\ and\ \bibinfo {author} {\bibfnamefont {F.}~\bibnamefont
  {Zamora}},\ }\href {https://doi.org/10.1039/c9ta06072c} {\bibfield  {journal}
  {\bibinfo  {journal} {Journal of Materials Chemistry A}\ }\textbf {\bibinfo
  {volume} {7}},\ \bibinfo {pages} {22475} (\bibinfo {year}
  {2019})}\BibitemShut {NoStop}%
\bibitem [{\citenamefont {Ji}\ \emph {et~al.}(2016)\citenamefont {Ji},
  \citenamefont {Song}, \citenamefont {Liu}, \citenamefont {Yan}, \citenamefont
  {Huo}, \citenamefont {Zhang}, \citenamefont {Su}, \citenamefont {Liao},
  \citenamefont {Wang}, \citenamefont {Ni}, \citenamefont {Hao},\ and\
  \citenamefont {Zeng}}]{jiTwodimensionalAntimoneneSingle2016}%
  \BibitemOpen
  \bibfield  {author} {\bibinfo {author} {\bibfnamefont {J.}~\bibnamefont
  {Ji}}, \bibinfo {author} {\bibfnamefont {X.}~\bibnamefont {Song}}, \bibinfo
  {author} {\bibfnamefont {J.}~\bibnamefont {Liu}}, \bibinfo {author}
  {\bibfnamefont {Z.}~\bibnamefont {Yan}}, \bibinfo {author} {\bibfnamefont
  {C.}~\bibnamefont {Huo}}, \bibinfo {author} {\bibfnamefont {S.}~\bibnamefont
  {Zhang}}, \bibinfo {author} {\bibfnamefont {M.}~\bibnamefont {Su}}, \bibinfo
  {author} {\bibfnamefont {L.}~\bibnamefont {Liao}}, \bibinfo {author}
  {\bibfnamefont {W.}~\bibnamefont {Wang}}, \bibinfo {author} {\bibfnamefont
  {Z.}~\bibnamefont {Ni}}, \bibinfo {author} {\bibfnamefont {Y.}~\bibnamefont
  {Hao}},\ and\ \bibinfo {author} {\bibfnamefont {H.}~\bibnamefont {Zeng}},\
  }\href {https://doi.org/10.1038/ncomms13352} {\bibfield  {journal} {\bibinfo
  {journal} {Nature Communications}\ }\textbf {\bibinfo {volume} {7}},\
  \bibinfo {pages} {13352} (\bibinfo {year} {2016})}\BibitemShut {NoStop}%
\bibitem [{\citenamefont {Wu}\ \emph {et~al.}(2017)\citenamefont {Wu},
  \citenamefont {Shao}, \citenamefont {Liu}, \citenamefont {Feng},
  \citenamefont {Wang}, \citenamefont {Sun}, \citenamefont {Liu}, \citenamefont
  {Wang}, \citenamefont {Liu}, \citenamefont {Zhu}, \citenamefont {Wang},
  \citenamefont {Du}, \citenamefont {Shi}, \citenamefont {Ibrahim},\ and\
  \citenamefont {Gao}}]{wuEpitaxialGrowthAirStability2017}%
  \BibitemOpen
  \bibfield  {author} {\bibinfo {author} {\bibfnamefont {X.}~\bibnamefont
  {Wu}}, \bibinfo {author} {\bibfnamefont {Y.}~\bibnamefont {Shao}}, \bibinfo
  {author} {\bibfnamefont {H.}~\bibnamefont {Liu}}, \bibinfo {author}
  {\bibfnamefont {Z.}~\bibnamefont {Feng}}, \bibinfo {author} {\bibfnamefont
  {Y.-L.}\ \bibnamefont {Wang}}, \bibinfo {author} {\bibfnamefont {J.-T.}\
  \bibnamefont {Sun}}, \bibinfo {author} {\bibfnamefont {C.}~\bibnamefont
  {Liu}}, \bibinfo {author} {\bibfnamefont {J.-O.}\ \bibnamefont {Wang}},
  \bibinfo {author} {\bibfnamefont {Z.-L.}\ \bibnamefont {Liu}}, \bibinfo
  {author} {\bibfnamefont {S.-Y.}\ \bibnamefont {Zhu}}, \bibinfo {author}
  {\bibfnamefont {Y.-Q.}\ \bibnamefont {Wang}}, \bibinfo {author}
  {\bibfnamefont {S.-X.}\ \bibnamefont {Du}}, \bibinfo {author} {\bibfnamefont
  {Y.-G.}\ \bibnamefont {Shi}}, \bibinfo {author} {\bibfnamefont
  {K.}~\bibnamefont {Ibrahim}},\ and\ \bibinfo {author} {\bibfnamefont {H.-J.}\
  \bibnamefont {Gao}},\ }\href {https://doi.org/10.1002/adma.201605407}
  {\bibfield  {journal} {\bibinfo  {journal} {Advanced Materials}\ }\textbf
  {\bibinfo {volume} {29}},\ \bibinfo {pages} {1605407} (\bibinfo {year}
  {2017})}\BibitemShut {NoStop}%
\bibitem [{\citenamefont {Chen}\ \emph {et~al.}(2018)\citenamefont {Chen},
  \citenamefont {Sun}, \citenamefont {Wu}, \citenamefont {Wang}, \citenamefont
  {Lee}, \citenamefont {Pao},\ and\ \citenamefont
  {Lin}}]{chenSingleCrystalAntimoneneFilms2018}%
  \BibitemOpen
  \bibfield  {author} {\bibinfo {author} {\bibfnamefont {H.-A.}\ \bibnamefont
  {Chen}}, \bibinfo {author} {\bibfnamefont {H.}~\bibnamefont {Sun}}, \bibinfo
  {author} {\bibfnamefont {C.-R.}\ \bibnamefont {Wu}}, \bibinfo {author}
  {\bibfnamefont {Y.-X.}\ \bibnamefont {Wang}}, \bibinfo {author}
  {\bibfnamefont {P.-H.}\ \bibnamefont {Lee}}, \bibinfo {author} {\bibfnamefont
  {C.-W.}\ \bibnamefont {Pao}},\ and\ \bibinfo {author} {\bibfnamefont {S.-Y.}\
  \bibnamefont {Lin}},\ }\href {https://doi.org/10.1021/acsami.8b02394}
  {\bibfield  {journal} {\bibinfo  {journal} {ACS Applied Materials \&
  Interfaces}\ }\textbf {\bibinfo {volume} {10}},\ \bibinfo {pages} {15058}
  (\bibinfo {year} {2018})}\BibitemShut {NoStop}%
\bibitem [{\citenamefont {Sun}\ \emph {et~al.}(2018)\citenamefont {Sun},
  \citenamefont {Lu}, \citenamefont {Xiang}, \citenamefont {Wang},
  \citenamefont {Shi}, \citenamefont {Wang}, \citenamefont {Washington},\ and\
  \citenamefont {Lu}}]{sunVanWaalsEpitaxy2018}%
  \BibitemOpen
  \bibfield  {author} {\bibinfo {author} {\bibfnamefont {X.}~\bibnamefont
  {Sun}}, \bibinfo {author} {\bibfnamefont {Z.}~\bibnamefont {Lu}}, \bibinfo
  {author} {\bibfnamefont {Y.}~\bibnamefont {Xiang}}, \bibinfo {author}
  {\bibfnamefont {Y.}~\bibnamefont {Wang}}, \bibinfo {author} {\bibfnamefont
  {J.}~\bibnamefont {Shi}}, \bibinfo {author} {\bibfnamefont {G.-C.}\
  \bibnamefont {Wang}}, \bibinfo {author} {\bibfnamefont {M.~A.}\ \bibnamefont
  {Washington}},\ and\ \bibinfo {author} {\bibfnamefont {T.-M.}\ \bibnamefont
  {Lu}},\ }\href {https://doi.org/10.1021/acsnano.8b02374} {\bibfield
  {journal} {\bibinfo  {journal} {ACS Nano}\ }\textbf {\bibinfo {volume}
  {12}},\ \bibinfo {pages} {6100} (\bibinfo {year} {2018})}\BibitemShut
  {NoStop}%
\bibitem [{\citenamefont {Flammini}\ \emph {et~al.}(2018)\citenamefont
  {Flammini}, \citenamefont {Colonna}, \citenamefont {Hogan}, \citenamefont
  {Mahatha}, \citenamefont {Papagno}, \citenamefont {Barla}, \citenamefont
  {Sheverdyaeva}, \citenamefont {Moras}, \citenamefont {Aliev}, \citenamefont
  {Babanly}, \citenamefont {Chulkov}, \citenamefont {Carbone},\ and\
  \citenamefont {Ronci}}]{flamminiEvidenceBetaAntimonene2018}%
  \BibitemOpen
  \bibfield  {author} {\bibinfo {author} {\bibfnamefont {R.}~\bibnamefont
  {Flammini}}, \bibinfo {author} {\bibfnamefont {S.}~\bibnamefont {Colonna}},
  \bibinfo {author} {\bibfnamefont {C.}~\bibnamefont {Hogan}}, \bibinfo
  {author} {\bibfnamefont {S.~K.}\ \bibnamefont {Mahatha}}, \bibinfo {author}
  {\bibfnamefont {M.}~\bibnamefont {Papagno}}, \bibinfo {author} {\bibfnamefont
  {A.}~\bibnamefont {Barla}}, \bibinfo {author} {\bibfnamefont {P.~M.}\
  \bibnamefont {Sheverdyaeva}}, \bibinfo {author} {\bibfnamefont
  {P.}~\bibnamefont {Moras}}, \bibinfo {author} {\bibfnamefont {Z.~S.}\
  \bibnamefont {Aliev}}, \bibinfo {author} {\bibfnamefont {M.~B.}\ \bibnamefont
  {Babanly}}, \bibinfo {author} {\bibfnamefont {E.~V.}\ \bibnamefont
  {Chulkov}}, \bibinfo {author} {\bibfnamefont {C.}~\bibnamefont {Carbone}},\
  and\ \bibinfo {author} {\bibfnamefont {F.}~\bibnamefont {Ronci}},\ }\href
  {https://doi.org/10.1088/1361-6528/aaa2c4} {\bibfield  {journal} {\bibinfo
  {journal} {Nanotechnology}\ }\textbf {\bibinfo {volume} {29}},\ \bibinfo
  {pages} {065704} (\bibinfo {year} {2018})}\BibitemShut {NoStop}%
\bibitem [{\citenamefont {Zhang}\ \emph {et~al.}(2020)\citenamefont {Zhang},
  \citenamefont {Li}, \citenamefont {Wu}, \citenamefont {Chang}, \citenamefont
  {Chang}, \citenamefont {Shih},\ and\ \citenamefont
  {Lin}}]{zhangTungstenDiselenideTopgate2020}%
  \BibitemOpen
  \bibfield  {author} {\bibinfo {author} {\bibfnamefont {Y.-W.}\ \bibnamefont
  {Zhang}}, \bibinfo {author} {\bibfnamefont {J.-Y.}\ \bibnamefont {Li}},
  \bibinfo {author} {\bibfnamefont {C.-H.}\ \bibnamefont {Wu}}, \bibinfo
  {author} {\bibfnamefont {C.-Y.}\ \bibnamefont {Chang}}, \bibinfo {author}
  {\bibfnamefont {S.-W.}\ \bibnamefont {Chang}}, \bibinfo {author}
  {\bibfnamefont {M.-H.}\ \bibnamefont {Shih}},\ and\ \bibinfo {author}
  {\bibfnamefont {S.-Y.}\ \bibnamefont {Lin}},\ }\href
  {https://doi.org/10.1038/s41598-020-63098-1} {\bibfield  {journal} {\bibinfo
  {journal} {Scientific Reports}\ }\textbf {\bibinfo {volume} {10}},\ \bibinfo
  {pages} {5967} (\bibinfo {year} {2020})}\BibitemShut {NoStop}%
\bibitem [{\citenamefont {Mao}\ \emph {et~al.}(2018)\citenamefont {Mao},
  \citenamefont {Zhang}, \citenamefont {Wang}, \citenamefont {Shan},
  \citenamefont {Zhai}, \citenamefont {Hu}, \citenamefont {Zhao},\ and\
  \citenamefont {Wang}}]{maoEpitaxialGrowthHighly2018}%
  \BibitemOpen
  \bibfield  {author} {\bibinfo {author} {\bibfnamefont {Y.-H.}\ \bibnamefont
  {Mao}}, \bibinfo {author} {\bibfnamefont {L.-F.}\ \bibnamefont {Zhang}},
  \bibinfo {author} {\bibfnamefont {H.-L.}\ \bibnamefont {Wang}}, \bibinfo
  {author} {\bibfnamefont {H.}~\bibnamefont {Shan}}, \bibinfo {author}
  {\bibfnamefont {X.-F.}\ \bibnamefont {Zhai}}, \bibinfo {author}
  {\bibfnamefont {Z.-P.}\ \bibnamefont {Hu}}, \bibinfo {author} {\bibfnamefont
  {A.-D.}\ \bibnamefont {Zhao}},\ and\ \bibinfo {author} {\bibfnamefont
  {B.}~\bibnamefont {Wang}},\ }\href
  {https://doi.org/10.1007/s11467-018-0757-3} {\bibfield  {journal} {\bibinfo
  {journal} {Frontiers of Physics}\ }\textbf {\bibinfo {volume} {13}},\
  \bibinfo {pages} {138106} (\bibinfo {year} {2018})}\BibitemShut {NoStop}%
\bibitem [{\citenamefont {Shao}\ \emph {et~al.}(2018)\citenamefont {Shao},
  \citenamefont {Liu}, \citenamefont {Cheng}, \citenamefont {Wu}, \citenamefont
  {Liu}, \citenamefont {Liu}, \citenamefont {Wang}, \citenamefont {Zhu},
  \citenamefont {Wang}, \citenamefont {Shi}, \citenamefont {Ibrahim},
  \citenamefont {Sun}, \citenamefont {Wang},\ and\ \citenamefont
  {Gao}}]{shaoEpitaxialGrowthFlat2018}%
  \BibitemOpen
  \bibfield  {author} {\bibinfo {author} {\bibfnamefont {Y.}~\bibnamefont
  {Shao}}, \bibinfo {author} {\bibfnamefont {Z.-L.}\ \bibnamefont {Liu}},
  \bibinfo {author} {\bibfnamefont {C.}~\bibnamefont {Cheng}}, \bibinfo
  {author} {\bibfnamefont {X.}~\bibnamefont {Wu}}, \bibinfo {author}
  {\bibfnamefont {H.}~\bibnamefont {Liu}}, \bibinfo {author} {\bibfnamefont
  {C.}~\bibnamefont {Liu}}, \bibinfo {author} {\bibfnamefont {J.-O.}\
  \bibnamefont {Wang}}, \bibinfo {author} {\bibfnamefont {S.-Y.}\ \bibnamefont
  {Zhu}}, \bibinfo {author} {\bibfnamefont {Y.-Q.}\ \bibnamefont {Wang}},
  \bibinfo {author} {\bibfnamefont {D.-X.}\ \bibnamefont {Shi}}, \bibinfo
  {author} {\bibfnamefont {K.}~\bibnamefont {Ibrahim}}, \bibinfo {author}
  {\bibfnamefont {J.-T.}\ \bibnamefont {Sun}}, \bibinfo {author} {\bibfnamefont
  {Y.-L.}\ \bibnamefont {Wang}},\ and\ \bibinfo {author} {\bibfnamefont
  {H.-J.}\ \bibnamefont {Gao}},\ }\href
  {https://doi.org/10.1021/acs.nanolett.8b00429} {\bibfield  {journal}
  {\bibinfo  {journal} {Nano Letters}\ }\textbf {\bibinfo {volume} {18}},\
  \bibinfo {pages} {2133} (\bibinfo {year} {2018})}\BibitemShut {NoStop}%
\bibitem [{\citenamefont {Sun}\ \emph {et~al.}(2020)\citenamefont {Sun},
  \citenamefont {Yang}, \citenamefont {Luo}, \citenamefont {Gou}, \citenamefont
  {Huang}, \citenamefont {Gu}, \citenamefont {Ma}, \citenamefont {Lian},
  \citenamefont {Duan}, \citenamefont {Wee}, \citenamefont {Lai}, \citenamefont
  {Zhang}, \citenamefont {Feng},\ and\ \citenamefont
  {Chen}}]{sunRealizationBuckledAntimonene2020}%
  \BibitemOpen
  \bibfield  {author} {\bibinfo {author} {\bibfnamefont {S.}~\bibnamefont
  {Sun}}, \bibinfo {author} {\bibfnamefont {T.}~\bibnamefont {Yang}}, \bibinfo
  {author} {\bibfnamefont {Y.~Z.}\ \bibnamefont {Luo}}, \bibinfo {author}
  {\bibfnamefont {J.}~\bibnamefont {Gou}}, \bibinfo {author} {\bibfnamefont
  {Y.}~\bibnamefont {Huang}}, \bibinfo {author} {\bibfnamefont
  {C.}~\bibnamefont {Gu}}, \bibinfo {author} {\bibfnamefont {Z.}~\bibnamefont
  {Ma}}, \bibinfo {author} {\bibfnamefont {X.}~\bibnamefont {Lian}}, \bibinfo
  {author} {\bibfnamefont {S.}~\bibnamefont {Duan}}, \bibinfo {author}
  {\bibfnamefont {A.~T.~S.}\ \bibnamefont {Wee}}, \bibinfo {author}
  {\bibfnamefont {M.}~\bibnamefont {Lai}}, \bibinfo {author} {\bibfnamefont
  {J.~L.}\ \bibnamefont {Zhang}}, \bibinfo {author} {\bibfnamefont {Y.~P.}\
  \bibnamefont {Feng}},\ and\ \bibinfo {author} {\bibfnamefont
  {W.}~\bibnamefont {Chen}},\ }\href
  {https://doi.org/10.1021/acs.jpclett.0c02637} {\bibfield  {journal} {\bibinfo
   {journal} {The Journal of Physical Chemistry Letters}\ }\textbf {\bibinfo
  {volume} {11}},\ \bibinfo {pages} {8976} (\bibinfo {year}
  {2020})}\BibitemShut {NoStop}%
\bibitem [{\citenamefont {Niu}\ \emph {et~al.}(2019)\citenamefont {Niu},
  \citenamefont {Zhou}, \citenamefont {Zhou}, \citenamefont {Hu}, \citenamefont
  {Zhang}, \citenamefont {Zhang}, \citenamefont {Zhou}, \citenamefont {Fuchs},\
  and\ \citenamefont {Zeng}}]{niuModulatingEpitaxialAtomic2019}%
  \BibitemOpen
  \bibfield  {author} {\bibinfo {author} {\bibfnamefont {T.}~\bibnamefont
  {Niu}}, \bibinfo {author} {\bibfnamefont {W.}~\bibnamefont {Zhou}}, \bibinfo
  {author} {\bibfnamefont {D.}~\bibnamefont {Zhou}}, \bibinfo {author}
  {\bibfnamefont {X.}~\bibnamefont {Hu}}, \bibinfo {author} {\bibfnamefont
  {S.}~\bibnamefont {Zhang}}, \bibinfo {author} {\bibfnamefont
  {K.}~\bibnamefont {Zhang}}, \bibinfo {author} {\bibfnamefont
  {M.}~\bibnamefont {Zhou}}, \bibinfo {author} {\bibfnamefont {H.}~\bibnamefont
  {Fuchs}},\ and\ \bibinfo {author} {\bibfnamefont {H.}~\bibnamefont {Zeng}},\
  }\href {https://doi.org/10.1002/adma.201902606} {\bibfield  {journal}
  {\bibinfo  {journal} {Advanced Materials}\ }\textbf {\bibinfo {volume}
  {31}},\ \bibinfo {pages} {1902606} (\bibinfo {year} {2019})}\BibitemShut
  {NoStop}%
\bibitem [{\citenamefont {Zhou}\ \emph {et~al.}(2019)\citenamefont {Zhou},
  \citenamefont {Si}, \citenamefont {Jiang}, \citenamefont {Song},
  \citenamefont {Huang}, \citenamefont {Ji},\ and\ \citenamefont
  {Niu}}]{zhouInterfacialEffectsGrowth2019}%
  \BibitemOpen
  \bibfield  {author} {\bibinfo {author} {\bibfnamefont {D.}~\bibnamefont
  {Zhou}}, \bibinfo {author} {\bibfnamefont {N.}~\bibnamefont {Si}}, \bibinfo
  {author} {\bibfnamefont {B.}~\bibnamefont {Jiang}}, \bibinfo {author}
  {\bibfnamefont {X.}~\bibnamefont {Song}}, \bibinfo {author} {\bibfnamefont
  {H.}~\bibnamefont {Huang}}, \bibinfo {author} {\bibfnamefont
  {Q.}~\bibnamefont {Ji}},\ and\ \bibinfo {author} {\bibfnamefont
  {T.}~\bibnamefont {Niu}},\ }\href {https://doi.org/10.1002/admi.201901050}
  {\bibfield  {journal} {\bibinfo  {journal} {Advanced Materials Interfaces}\
  }\textbf {\bibinfo {volume} {6}},\ \bibinfo {pages} {1901050} (\bibinfo
  {year} {2019})}\BibitemShut {NoStop}%
\bibitem [{\citenamefont {Niu}\ \emph {et~al.}(2020)\citenamefont {Niu},
  \citenamefont {Meng}, \citenamefont {Zhou}, \citenamefont {Si}, \citenamefont
  {Zhai}, \citenamefont {Hao}, \citenamefont {Zhou},\ and\ \citenamefont
  {Fuchs}}]{niuLargeScaleSynthesis2020}%
  \BibitemOpen
  \bibfield  {author} {\bibinfo {author} {\bibfnamefont {T.}~\bibnamefont
  {Niu}}, \bibinfo {author} {\bibfnamefont {Q.}~\bibnamefont {Meng}}, \bibinfo
  {author} {\bibfnamefont {D.}~\bibnamefont {Zhou}}, \bibinfo {author}
  {\bibfnamefont {N.}~\bibnamefont {Si}}, \bibinfo {author} {\bibfnamefont
  {S.}~\bibnamefont {Zhai}}, \bibinfo {author} {\bibfnamefont {X.}~\bibnamefont
  {Hao}}, \bibinfo {author} {\bibfnamefont {M.}~\bibnamefont {Zhou}},\ and\
  \bibinfo {author} {\bibfnamefont {H.}~\bibnamefont {Fuchs}},\ }\href
  {https://doi.org/10.1002/adma.201906873} {\bibfield  {journal} {\bibinfo
  {journal} {Advanced Materials}\ }\textbf {\bibinfo {volume} {32}},\ \bibinfo
  {pages} {1906873} (\bibinfo {year} {2020})}\BibitemShut {NoStop}%
\bibitem [{\citenamefont {Gu}\ \emph {et~al.}(2020)\citenamefont {Gu},
  \citenamefont {Li}, \citenamefont {Ding}, \citenamefont {Zhang},
  \citenamefont {Xia}, \citenamefont {Wang}, \citenamefont {Lu}, \citenamefont
  {Lu},\ and\ \citenamefont {Chen}}]{guDirectGrowthAntimonene2020}%
  \BibitemOpen
  \bibfield  {author} {\bibinfo {author} {\bibfnamefont {M.}~\bibnamefont
  {Gu}}, \bibinfo {author} {\bibfnamefont {C.}~\bibnamefont {Li}}, \bibinfo
  {author} {\bibfnamefont {Y.}~\bibnamefont {Ding}}, \bibinfo {author}
  {\bibfnamefont {K.}~\bibnamefont {Zhang}}, \bibinfo {author} {\bibfnamefont
  {S.}~\bibnamefont {Xia}}, \bibinfo {author} {\bibfnamefont {Y.}~\bibnamefont
  {Wang}}, \bibinfo {author} {\bibfnamefont {M.-H.}\ \bibnamefont {Lu}},
  \bibinfo {author} {\bibfnamefont {H.}~\bibnamefont {Lu}},\ and\ \bibinfo
  {author} {\bibfnamefont {Y.-F.}\ \bibnamefont {Chen}},\ }\href
  {https://doi.org/10.3390/app10020639} {\bibfield  {journal} {\bibinfo
  {journal} {Applied Sciences}\ }\textbf {\bibinfo {volume} {10}},\ \bibinfo
  {pages} {639} (\bibinfo {year} {2020})}\BibitemShut {NoStop}%
\bibitem [{\citenamefont {Wang}\ \emph {et~al.}(2010)\citenamefont {Wang},
  \citenamefont {Lv}, \citenamefont {Zhu},\ and\ \citenamefont
  {Ma}}]{wangCrystalStructurePrediction2010}%
  \BibitemOpen
  \bibfield  {author} {\bibinfo {author} {\bibfnamefont {Y.}~\bibnamefont
  {Wang}}, \bibinfo {author} {\bibfnamefont {J.}~\bibnamefont {Lv}}, \bibinfo
  {author} {\bibfnamefont {L.}~\bibnamefont {Zhu}},\ and\ \bibinfo {author}
  {\bibfnamefont {Y.}~\bibnamefont {Ma}},\ }\href
  {https://doi.org/10.1103/PhysRevB.82.094116} {\bibfield  {journal} {\bibinfo
  {journal} {Physical Review B}\ }\textbf {\bibinfo {volume} {82}},\ \bibinfo
  {pages} {094116} (\bibinfo {year} {2010})}\BibitemShut {NoStop}%
\bibitem [{\citenamefont {Wang}\ \emph {et~al.}(2012)\citenamefont {Wang},
  \citenamefont {Lv}, \citenamefont {Zhu},\ and\ \citenamefont
  {Ma}}]{wangCALYPSOMethodCrystal2012}%
  \BibitemOpen
  \bibfield  {author} {\bibinfo {author} {\bibfnamefont {Y.}~\bibnamefont
  {Wang}}, \bibinfo {author} {\bibfnamefont {J.}~\bibnamefont {Lv}}, \bibinfo
  {author} {\bibfnamefont {L.}~\bibnamefont {Zhu}},\ and\ \bibinfo {author}
  {\bibfnamefont {Y.}~\bibnamefont {Ma}},\ }\href
  {https://doi.org/10.1016/j.cpc.2012.05.008} {\bibfield  {journal} {\bibinfo
  {journal} {Computer Physics Communications}\ }\textbf {\bibinfo {volume}
  {183}},\ \bibinfo {pages} {2063} (\bibinfo {year} {2012})}\BibitemShut
  {NoStop}%
\bibitem [{\citenamefont {Lu}\ \emph {et~al.}(2014)\citenamefont {Lu},
  \citenamefont {Wang}, \citenamefont {Liu}, \citenamefont {Miao},\ and\
  \citenamefont {Ma}}]{luSelfassembledUltrathinNanotubes2014}%
  \BibitemOpen
  \bibfield  {author} {\bibinfo {author} {\bibfnamefont {S.}~\bibnamefont
  {Lu}}, \bibinfo {author} {\bibfnamefont {Y.}~\bibnamefont {Wang}}, \bibinfo
  {author} {\bibfnamefont {H.}~\bibnamefont {Liu}}, \bibinfo {author}
  {\bibfnamefont {M.-s.}\ \bibnamefont {Miao}},\ and\ \bibinfo {author}
  {\bibfnamefont {Y.}~\bibnamefont {Ma}},\ }\href
  {https://doi.org/10.1038/ncomms4666} {\bibfield  {journal} {\bibinfo
  {journal} {Nature Communications}\ }\textbf {\bibinfo {volume} {5}},\
  \bibinfo {pages} {3666} (\bibinfo {year} {2014})}\BibitemShut {NoStop}%
\bibitem [{\citenamefont {Oppo}\ \emph {et~al.}(1993)\citenamefont {Oppo},
  \citenamefont {Fiorentini},\ and\ \citenamefont
  {Scheffler}}]{oppoTheoryAdsorptionSurfactant1993}%
  \BibitemOpen
  \bibfield  {author} {\bibinfo {author} {\bibfnamefont {S.}~\bibnamefont
  {Oppo}}, \bibinfo {author} {\bibfnamefont {V.}~\bibnamefont {Fiorentini}},\
  and\ \bibinfo {author} {\bibfnamefont {M.}~\bibnamefont {Scheffler}},\ }\href
  {https://doi.org/10.1103/PhysRevLett.71.2437} {\bibfield  {journal} {\bibinfo
   {journal} {Physical Review Letters}\ }\textbf {\bibinfo {volume} {71}},\
  \bibinfo {pages} {2437} (\bibinfo {year} {1993})}\BibitemShut {NoStop}%
\bibitem [{\citenamefont {Soares}\ \emph {et~al.}(2000)\citenamefont {Soares},
  \citenamefont {Bittencourt}, \citenamefont {Nascimento}, \citenamefont {{de
  Carvalho}}, \citenamefont {{de Castilho}}, \citenamefont {McConville},
  \citenamefont {{de Carvalho}},\ and\ \citenamefont
  {Woodruff}}]{soaresStructureDeterminationAg2000}%
  \BibitemOpen
  \bibfield  {author} {\bibinfo {author} {\bibfnamefont {E.~A.}\ \bibnamefont
  {Soares}}, \bibinfo {author} {\bibfnamefont {C.}~\bibnamefont {Bittencourt}},
  \bibinfo {author} {\bibfnamefont {V.~B.}\ \bibnamefont {Nascimento}},
  \bibinfo {author} {\bibfnamefont {V.~E.}\ \bibnamefont {{de Carvalho}}},
  \bibinfo {author} {\bibfnamefont {C.~M.~C.}\ \bibnamefont {{de Castilho}}},
  \bibinfo {author} {\bibfnamefont {C.~F.}\ \bibnamefont {McConville}},
  \bibinfo {author} {\bibfnamefont {A.~V.}\ \bibnamefont {{de Carvalho}}},\
  and\ \bibinfo {author} {\bibfnamefont {D.~P.}\ \bibnamefont {Woodruff}},\
  }\href {https://doi.org/10.1103/PhysRevB.61.13983} {\bibfield  {journal}
  {\bibinfo  {journal} {Physical Review B}\ }\textbf {\bibinfo {volume} {61}},\
  \bibinfo {pages} {13983} (\bibinfo {year} {2000})}\BibitemShut {NoStop}%
\bibitem [{\citenamefont {Quinn}\ \emph {et~al.}(2002)\citenamefont {Quinn},
  \citenamefont {Brown}, \citenamefont {Woodruff}, \citenamefont {Bailey},\
  and\ \citenamefont {Noakes}}]{quinnStructuralStudyAdsorption2002}%
  \BibitemOpen
  \bibfield  {author} {\bibinfo {author} {\bibfnamefont {P.~D.}\ \bibnamefont
  {Quinn}}, \bibinfo {author} {\bibfnamefont {D.}~\bibnamefont {Brown}},
  \bibinfo {author} {\bibfnamefont {D.~P.}\ \bibnamefont {Woodruff}}, \bibinfo
  {author} {\bibfnamefont {P.}~\bibnamefont {Bailey}},\ and\ \bibinfo {author}
  {\bibfnamefont {T.~C.~Q.}\ \bibnamefont {Noakes}},\ }\href
  {https://doi.org/10.1016/S0039-6028(02)01488-7} {\bibfield  {journal}
  {\bibinfo  {journal} {Surface Science}\ }\textbf {\bibinfo {volume} {511}},\
  \bibinfo {pages} {43} (\bibinfo {year} {2002})}\BibitemShut {NoStop}%
\bibitem [{\citenamefont {Kresse}\ and\ \citenamefont
  {Hafner}(1994)}]{kresseInitioMoleculardynamicsSimulation1994}%
  \BibitemOpen
  \bibfield  {author} {\bibinfo {author} {\bibfnamefont {G.}~\bibnamefont
  {Kresse}}\ and\ \bibinfo {author} {\bibfnamefont {J.}~\bibnamefont
  {Hafner}},\ }\href {https://doi.org/10.1103/PhysRevB.49.14251} {\bibfield
  {journal} {\bibinfo  {journal} {Physical Review B}\ }\textbf {\bibinfo
  {volume} {49}},\ \bibinfo {pages} {14251} (\bibinfo {year}
  {1994})}\BibitemShut {NoStop}%
\bibitem [{\citenamefont {Kresse}\ and\ \citenamefont
  {Furthm{\"u}ller}(1996{\natexlab{a}})}]{kresseEfficiencyAbinitioTotal1996}%
  \BibitemOpen
  \bibfield  {author} {\bibinfo {author} {\bibfnamefont {G.}~\bibnamefont
  {Kresse}}\ and\ \bibinfo {author} {\bibfnamefont {J.}~\bibnamefont
  {Furthm{\"u}ller}},\ }\href {https://doi.org/10.1016/0927-0256(96)00008-0}
  {\bibfield  {journal} {\bibinfo  {journal} {Computational Materials Science}\
  }\textbf {\bibinfo {volume} {6}},\ \bibinfo {pages} {15} (\bibinfo {year}
  {1996}{\natexlab{a}})}\BibitemShut {NoStop}%
\bibitem [{\citenamefont {Kresse}\ and\ \citenamefont
  {Furthm{\"u}ller}(1996{\natexlab{b}})}]{kresseEfficientIterativeSchemes1996}%
  \BibitemOpen
  \bibfield  {author} {\bibinfo {author} {\bibfnamefont {G.}~\bibnamefont
  {Kresse}}\ and\ \bibinfo {author} {\bibfnamefont {J.}~\bibnamefont
  {Furthm{\"u}ller}},\ }\href {https://doi.org/10.1103/PhysRevB.54.11169}
  {\bibfield  {journal} {\bibinfo  {journal} {Physical Review B}\ }\textbf
  {\bibinfo {volume} {54}},\ \bibinfo {pages} {11169} (\bibinfo {year}
  {1996}{\natexlab{b}})}\BibitemShut {NoStop}%
\bibitem [{\citenamefont
  {Bl{\"o}chl}(1994)}]{blochlProjectorAugmentedwaveMethod1994}%
  \BibitemOpen
  \bibfield  {author} {\bibinfo {author} {\bibfnamefont {P.~E.}\ \bibnamefont
  {Bl{\"o}chl}},\ }\href {https://doi.org/10.1103/PhysRevB.50.17953} {\bibfield
   {journal} {\bibinfo  {journal} {Physical Review B}\ }\textbf {\bibinfo
  {volume} {50}},\ \bibinfo {pages} {17953} (\bibinfo {year}
  {1994})}\BibitemShut {NoStop}%
\bibitem [{\citenamefont {Perdew}\ \emph {et~al.}(1992)\citenamefont {Perdew},
  \citenamefont {Chevary}, \citenamefont {Vosko}, \citenamefont {Jackson},
  \citenamefont {Pederson}, \citenamefont {Singh},\ and\ \citenamefont
  {Fiolhais}}]{perdewAtomsMoleculesSolids1992}%
  \BibitemOpen
  \bibfield  {author} {\bibinfo {author} {\bibfnamefont {J.~P.}\ \bibnamefont
  {Perdew}}, \bibinfo {author} {\bibfnamefont {J.~A.}\ \bibnamefont {Chevary}},
  \bibinfo {author} {\bibfnamefont {S.~H.}\ \bibnamefont {Vosko}}, \bibinfo
  {author} {\bibfnamefont {K.~A.}\ \bibnamefont {Jackson}}, \bibinfo {author}
  {\bibfnamefont {M.~R.}\ \bibnamefont {Pederson}}, \bibinfo {author}
  {\bibfnamefont {D.~J.}\ \bibnamefont {Singh}},\ and\ \bibinfo {author}
  {\bibfnamefont {C.}~\bibnamefont {Fiolhais}},\ }\href
  {https://doi.org/10.1103/PhysRevB.46.6671} {\bibfield  {journal} {\bibinfo
  {journal} {Physical Review B}\ }\textbf {\bibinfo {volume} {46}},\ \bibinfo
  {pages} {6671} (\bibinfo {year} {1992})}\BibitemShut {NoStop}%
\bibitem [{\citenamefont {Perdew}\ \emph {et~al.}(1996)\citenamefont {Perdew},
  \citenamefont {Burke},\ and\ \citenamefont
  {Ernzerhof}}]{perdewGeneralizedGradientApproximation1996}%
  \BibitemOpen
  \bibfield  {author} {\bibinfo {author} {\bibfnamefont {J.~P.}\ \bibnamefont
  {Perdew}}, \bibinfo {author} {\bibfnamefont {K.}~\bibnamefont {Burke}},\ and\
  \bibinfo {author} {\bibfnamefont {M.}~\bibnamefont {Ernzerhof}},\ }\href
  {https://doi.org/10.1103/PhysRevLett.77.3865} {\bibfield  {journal} {\bibinfo
   {journal} {Physical Review Letters}\ }\textbf {\bibinfo {volume} {77}},\
  \bibinfo {pages} {3865} (\bibinfo {year} {1996})}\BibitemShut {NoStop}%
\bibitem [{\citenamefont {Grimme}\ \emph {et~al.}(2010)\citenamefont {Grimme},
  \citenamefont {Antony}, \citenamefont {Ehrlich},\ and\ \citenamefont
  {Krieg}}]{grimmeConsistentAccurateInitio2010}%
  \BibitemOpen
  \bibfield  {author} {\bibinfo {author} {\bibfnamefont {S.}~\bibnamefont
  {Grimme}}, \bibinfo {author} {\bibfnamefont {J.}~\bibnamefont {Antony}},
  \bibinfo {author} {\bibfnamefont {S.}~\bibnamefont {Ehrlich}},\ and\ \bibinfo
  {author} {\bibfnamefont {H.}~\bibnamefont {Krieg}},\ }\href
  {https://doi.org/10.1063/1.3382344} {\bibfield  {journal} {\bibinfo
  {journal} {The Journal of Chemical Physics}\ }\textbf {\bibinfo {volume}
  {132}},\ \bibinfo {pages} {154104} (\bibinfo {year} {2010})}\BibitemShut
  {NoStop}%
\bibitem [{\citenamefont {K{\"u}hne}\ \emph {et~al.}(2020)\citenamefont
  {K{\"u}hne}, \citenamefont {Iannuzzi}, \citenamefont {Del~Ben}, \citenamefont
  {Rybkin}, \citenamefont {Seewald}, \citenamefont {Stein}, \citenamefont
  {Laino}, \citenamefont {Khaliullin}, \citenamefont {Sch{\"u}tt},
  \citenamefont {Schiffmann}, \citenamefont {Golze}, \citenamefont {Wilhelm},
  \citenamefont {Chulkov}, \citenamefont {{Bani-Hashemian}}, \citenamefont
  {Weber}, \citenamefont {Bor{\v s}tnik}, \citenamefont {Taillefumier},
  \citenamefont {Jakobovits}, \citenamefont {Lazzaro}, \citenamefont {Pabst},
  \citenamefont {M{\"u}ller}, \citenamefont {Schade}, \citenamefont {Guidon},
  \citenamefont {Andermatt}, \citenamefont {Holmberg}, \citenamefont
  {Schenter}, \citenamefont {Hehn}, \citenamefont {Bussy}, \citenamefont
  {Belleflamme}, \citenamefont {Tabacchi}, \citenamefont {Gl{\"o}{\ss}},
  \citenamefont {Lass}, \citenamefont {Bethune}, \citenamefont {Mundy},
  \citenamefont {Plessl}, \citenamefont {Watkins}, \citenamefont
  {VandeVondele}, \citenamefont {Krack},\ and\ \citenamefont
  {Hutter}}]{kuhneCP2KElectronicStructure2020}%
  \BibitemOpen
  \bibfield  {author} {\bibinfo {author} {\bibfnamefont {T.~D.}\ \bibnamefont
  {K{\"u}hne}}, \bibinfo {author} {\bibfnamefont {M.}~\bibnamefont {Iannuzzi}},
  \bibinfo {author} {\bibfnamefont {M.}~\bibnamefont {Del~Ben}}, \bibinfo
  {author} {\bibfnamefont {V.~V.}\ \bibnamefont {Rybkin}}, \bibinfo {author}
  {\bibfnamefont {P.}~\bibnamefont {Seewald}}, \bibinfo {author} {\bibfnamefont
  {F.}~\bibnamefont {Stein}}, \bibinfo {author} {\bibfnamefont
  {T.}~\bibnamefont {Laino}}, \bibinfo {author} {\bibfnamefont {R.~Z.}\
  \bibnamefont {Khaliullin}}, \bibinfo {author} {\bibfnamefont
  {O.}~\bibnamefont {Sch{\"u}tt}}, \bibinfo {author} {\bibfnamefont
  {F.}~\bibnamefont {Schiffmann}}, \bibinfo {author} {\bibfnamefont
  {D.}~\bibnamefont {Golze}}, \bibinfo {author} {\bibfnamefont
  {J.}~\bibnamefont {Wilhelm}}, \bibinfo {author} {\bibfnamefont
  {S.}~\bibnamefont {Chulkov}}, \bibinfo {author} {\bibfnamefont {M.~H.}\
  \bibnamefont {{Bani-Hashemian}}}, \bibinfo {author} {\bibfnamefont
  {V.}~\bibnamefont {Weber}}, \bibinfo {author} {\bibfnamefont
  {U.}~\bibnamefont {Bor{\v s}tnik}}, \bibinfo {author} {\bibfnamefont
  {M.}~\bibnamefont {Taillefumier}}, \bibinfo {author} {\bibfnamefont {A.~S.}\
  \bibnamefont {Jakobovits}}, \bibinfo {author} {\bibfnamefont
  {A.}~\bibnamefont {Lazzaro}}, \bibinfo {author} {\bibfnamefont
  {H.}~\bibnamefont {Pabst}}, \bibinfo {author} {\bibfnamefont
  {T.}~\bibnamefont {M{\"u}ller}}, \bibinfo {author} {\bibfnamefont
  {R.}~\bibnamefont {Schade}}, \bibinfo {author} {\bibfnamefont
  {M.}~\bibnamefont {Guidon}}, \bibinfo {author} {\bibfnamefont
  {S.}~\bibnamefont {Andermatt}}, \bibinfo {author} {\bibfnamefont
  {N.}~\bibnamefont {Holmberg}}, \bibinfo {author} {\bibfnamefont {G.~K.}\
  \bibnamefont {Schenter}}, \bibinfo {author} {\bibfnamefont {A.}~\bibnamefont
  {Hehn}}, \bibinfo {author} {\bibfnamefont {A.}~\bibnamefont {Bussy}},
  \bibinfo {author} {\bibfnamefont {F.}~\bibnamefont {Belleflamme}}, \bibinfo
  {author} {\bibfnamefont {G.}~\bibnamefont {Tabacchi}}, \bibinfo {author}
  {\bibfnamefont {A.}~\bibnamefont {Gl{\"o}{\ss}}}, \bibinfo {author}
  {\bibfnamefont {M.}~\bibnamefont {Lass}}, \bibinfo {author} {\bibfnamefont
  {I.}~\bibnamefont {Bethune}}, \bibinfo {author} {\bibfnamefont {C.~J.}\
  \bibnamefont {Mundy}}, \bibinfo {author} {\bibfnamefont {C.}~\bibnamefont
  {Plessl}}, \bibinfo {author} {\bibfnamefont {M.}~\bibnamefont {Watkins}},
  \bibinfo {author} {\bibfnamefont {J.}~\bibnamefont {VandeVondele}}, \bibinfo
  {author} {\bibfnamefont {M.}~\bibnamefont {Krack}},\ and\ \bibinfo {author}
  {\bibfnamefont {J.}~\bibnamefont {Hutter}},\ }\href
  {https://doi.org/10.1063/5.0007045} {\bibfield  {journal} {\bibinfo
  {journal} {The Journal of Chemical Physics}\ }\textbf {\bibinfo {volume}
  {152}},\ \bibinfo {pages} {194103} (\bibinfo {year} {2020})}\BibitemShut
  {NoStop}%
\end{thebibliography}%

\clearpage
\pagestyle{empty}

\newpage

\newpage

\newpage

\newpage

\newpage

\newpage

\newpage

\newpage

\end{document}